

\def\gtorder{\mathrel{\raise.3ex\hbox{$>$}\mkern-14mu
             \lower0.6ex\hbox{$\sim$}}}
\def\ltorder{\mathrel{\raise.3ex\hbox{$<$}\mkern-14mu
             \lower0.6ex\hbox{$\sim$}}}

\def\page{\footline={\ifnum\pageno=1 \hfil
          \else\hss\twelverm\folio\hss\fi}}

\def\begintex{
\page\twelvepoint\rm\doublespace\raggedbottom
\begintitlepage
\def\endtitlepage{\eject}
\def\abstract{\centerline{\bf ABSTRACT}\smallskip}
\def\keywords##1\par{\parindent=.8in{\vskip.35in \item{\it
Keywords:} ##1\par} \parindent=20pt}
\def\section##1\par{\vfil\eject \centerline{\bf ##1}\par\medskip}
\def\sectiona##1\par{\vfil\eject \centerline{\bf ##1}\par}
\def\sectionb##1\par{\centerline{\bf ##1}\par\medskip}
\def\subsection##1\par{\bigbreak \centerline{\it 
##1}\par\smallskip}
\def\subsubsection##1\par{\bigbreak \noindent{\it ##1}\par}
\def\acknow{\vskip.4in\centerline{\bf 
Acknowledgements}\smallskip}
\def\references{\vfil\eject \centerline{\bf REFERENCES}\par
\bigskip\beginrefs}}

   \def\begintitlepage{\doublespace\null\vfil
   \def\title##1\par{\centerline{\bf##1}\par}
   \def\author##1\par{\vskip.4in{\centerline{\it ##1}}\par}
   \def\affil##1\par{\centerline{##1}\par}}

\def\ahead#1\smallskip{{\vfil\eject{\centerline{\bf#1}} 
\smallskip}{\message#1}}
\def\bhead#1\smallskip{{\bigbreak{\centerline{\it#1}}\smallskip} 
{\message#1}}
\def\chead#1\par{{\bigbreak\noindent{\it#1}\par} {\message#1}}

\def\head#1. #2\par{\medbreak\centerline{{\bf#1.\enspace}{\it#2}} 
\par\medbreak}

\def\levelone#1~ ~ #2\smallskip{\noindent#1~ ~ {\bf#2}\smallskip}
\def\leveltwo#1~ ~ #2\smallskip{\noindent#1~ ~ {\it#2}\smallskip}
\def\levelthree#1~ ~ #2\smallskip{\noindent#1~ ~ {#2}\smallskip}

\def\m{^m\kern-7pt .\kern+3.5pt}
\def\p{^{\prime\prime}\kern-2.1mm .\kern+.6mm}
\def\pone{^{\prime}\kern-1.05mm .\kern+.3mm}
\def\dpoint{^d\kern-1.05mm .\kern+.3mm}
\def\hpoint{^h\kern-2.1mm .\kern+.6mm}
\def\y{^y\kern-1.05mm .\kern+.3mm}
\def\s{^s\kern-1.2mm .\kern+.3mm}

\def\apgt{\ {\raise-.5ex\hbox{$\buildrel>\over\sim$}}\ }
\def\aplt{\ {\raise-.5ex\hbox{$\buildrel<\over\sim$}}\ }
\def\deg{^{\circ}}

\def\hup{^{h}\kern-2.1mm .\kern+.6mm}

%
%

%
%

%
%
\def\today{\number\day\space\ifcase\month\or
January\or February\or March\or April\or May\or June\or July\or 
August\or September\or October\or November\or December\fi
\space\number\year}

\def\sqr#1#2{{\vcenter{\hrule height.#2pt
\hbox{\vrule width.#2pt height#1pt \kern#1pt
\vrule width.#2pt}
\hrule height.#2pt}}}

\newcount\equationnumber
\newbox\eqprefix
\def\neweqprefix#1{\global\equationnumber=1
\global\setbox\eqprefix=\hbox{#1}}

\def\autono{(\copy\eqprefix\number\equationnumber)
\global\advance\equationnumber by 1}

%
%
%
%


\def\beginrefs{\begingroup\parindent=0pt\frenchspacing 
\parskip=1pt plus 1pt minus 1pt\interlinepenalty=1000 
\tolerance=400 \hyphenpenalty=10000
\everypar={\hangindent=1.6pc}

\def\nature##1,{{\it Nature}, {\bf##1},}

\def\aa##1,{{\it Astr.\ Ap.,\ }{\bf##1},}
\def\aapr{{\it Astr.\ Ap.,\ }in press.}
\def\ajaa##1,{{\it Astron.\ Astrophys.,\ }{\bf##1},}
\def\ajaapr{{\it Astron.\ Astrophys.,\ }in press.}

\def\aalet##1,{{\it Astr.\ Ap.\ (Letters),\ }{\bf##1},}
\def\aaletpr{{\it Astr.\ Ap.\ (Letters),\ }in press.}
\def\ajaalet##1,{{\it Astron. Astrophys. (Letters),\ }{\bf##1},}
\def\ajaaletpr{{\it Astron. Astrophys. (Letters),} in press.}

\def\aasup##1,{{\it Astr. Ap. Suppl.,\ }{\bf##1},}
\def\aasuppr{{\it Astr.\ Ap.\ Suppl.,\ }in press.}
\def\ajaasup##1,{{\it Astron. Astrophys. Suppl.,\ }{\bf##1},}
\def\ajaasuppr{{\it Astron.\ Astrophys.\ Suppl.,\ }in press.}

\def\aass##1,{{\it Astr. Ap. Suppl. Ser.,\ }{\bf##1},}
\def\aasspr{{\it Astr. Ap. Suppl. Ser.,} in press.}

\def\aj##1,{{\it A.~J.,\ }{\bf##1},}
\def\ajpr{{\it A.~J.,\ }in press.}
\def\ajaj##1,{{\it Astron.~J.,\ }{\bf##1},}
\def\ajajpr{{\it Astron.~J.,} in press.}

\def\apj##1,{{\it Ap.~J.,\  }{\bf##1},}
\def\apjpr{{\it Ap.~J.,} in press.}
\def\ajapj##1,{{\it Astrophys.~J.,\ }{\bf##1},}
\def\ajapjpr{{\it Astrophys.~J.,} in press.}

\def\apjlet##1,{{\it Ap.~J. (Letters),\ }{\bf##1},}
\def\apjletpr{{\it Ap.~J. (Letters),} in press.}
\def\ajapjlet##1,{{\it Astrophys. J. Lett.,\ }{\bf##1},}
\def\ajapjletpr{{\it Astrophys. J. Lett.,} in press.}

\def\apjsup##1,{{\it Ap.~J.~Suppl.,\ }{\bf##1},}
\def\apjsuppr{{\it Ap.~J.\ Suppl.,} in press.}
\def\ajapjsup##1,{{\it Astrophys. J. Suppl.,\ }{\bf##1},}
\def\ajapjsuppr{{\it Astrophys. J.\ Suppl.,} in press.}

\def\araa##1,{{\it Ann. Rev. A.~A.,\ }{\bf##1},}
\def\araapr {{\it Ann. Rev. A.~A.,} in press.}

\def\baas##1,{{\it B.A.A.S.,\ }{\bf##1},}
\def\baaspr{{\it B.A.A.S.,} in press.}

\def\mnras##1,{{\it M.N.R.A.S.,\ }{\bf##1},}
\def\mnraspr{{\it M.N.R.A.S.,} in press.}
\def\ajmnras##1,{{\it Mon. Not. R. Astron. Soc.,\ }{\bf##1},}
\def\ajmnraspr{{\it Mon. Not. R. Astron. Soc.,} in press.}

\def\pasp##1,{{\it Pub.~A.S.P.,\ }{\bf##1},}
\def\pasppr{{\it Pub.~A.S.P.,} in press.}
\def\ajpasp##1,{{\it Publ. Astron. Soc. Pac.,\ }{\bf##1},}
\def\ajpasppr{{\it Publ. Astron. Soc. Pac.,} in press.}
}
\def\endrefs{\endgroup}

\newcount\refno
\refno=0

\def\beginorefs\par{\begingroup\parindent=12pt
\frenchspacing \parskip=1pt plus 1pt minus 1pt
\interlinepenalty=1000 \tolerance=400 \hyphenpenalty=10000
\everypar={\item{\the\refno.}\hangindent=2.6pc}

\def\nature##1,{{\it Nature}, {\bf##1},}

\def\aa##1,{{\it Astr.\ Ap.,\ }{\bf##1},}
\def\aapr{{\it Astr.\ Ap.,\ }in press.}
\def\ajaa##1,{{\it Astron.\ Astrophys.,\ }{\bf##1},}
\def\ajaapr{{\it Astron.\ Astrophys.,\ }in press.}

\def\aalet##1,{{\it Astr.\ Ap.\ (Letters),\ }{\bf##1},}
\def\aaletpr{{\it Astr.\ Ap.\ (Letters),\ }in press.}
\def\ajaalet##1,{{\it Astron. Astrophys. (Letters),\ }{\bf##1},}
\def\ajaaletpr{{\it Astron. Astrophys. (Letters),} in press.}

\def\aasup##1,{{\it Astr. Ap. Suppl.,\ }{\bf##1},}
\def\aasuppr{{\it Astr.\ Ap.\ Suppl.,\ }in press.}
\def\ajaasup##1,{{\it Astron. Astrophys. Suppl.,\ }{\bf##1},}
\def\ajaasuppr{{\it Astron.\ Astrophys.\ Suppl.,\ }in press.}

\def\aass##1,{{\it Astr. Ap. Suppl. Ser.,\ }{\bf##1},}
\def\aasspr{{\it Astr. Ap. Suppl. Ser.,} in press.}

\def\aj##1,{{\it A.~J.,\ }{\bf##1},}
\def\ajpr{{\it A.~J.,\ }in press.}
\def\ajaj##1,{{\it Astron.~J.,\ }{\bf##1},}
\def\ajajpr{{\it Astron.~J.,} in press.}

\def\apj##1,{{\it Ap.~J.,\  }{\bf##1},}
\def\apjpr{{\it Ap.~J.,} in press.}
\def\ajapj##1,{{\it Astrophys.~J.,\ }{\bf##1},}
\def\ajapjpr{{\it Astrophys.~J.,} in press.}

\def\apjlet##1,{{\it Ap.~J. (Letters),\ }{\bf##1},}
\def\apjletpr{{\it Ap.~J. (Letters),} in press.}
\def\ajapjlet##1,{{\it Astrophys. J. Lett.,\ }{\bf##1},}
\def\ajapjletpr{{\it Astrophys. J. Lett.,} in press.}

\def\apjsup##1,{{\it Ap.~J.~Suppl.,\ }{\bf##1},}
\def\apjsuppr{{\it Ap.~J.\ Suppl.,} in press.}
\def\ajapjsup##1,{{\it Astrophys. J. Suppl.,\ }{\bf##1},}
\def\ajapjsuppr{{\it Astrophys. J.\ Suppl.,} in press.}

\def\araa##1,{{\it Ann. Rev. A.~A.,\ }{\bf##1},}
\def\araapr {{\it Ann. Rev. A.~A.,} in press.}

\def\baas##1,{{\it B.A.A.S.,\ }{\bf##1},}
\def\baaspr{{\it B.A.A.S.,} in press.}

\def\mnras##1,{{\it M.N.R.A.S.,\ }{\bf##1},}
\def\mnraspr{{\it M.N.R.A.S.,} in press.}
\def\ajmnras##1,{{\it Mon. Not. R. Astron. Soc.,\ }{\bf##1},}
\def\ajmnraspr{{\it Mon. Not. R. Astron. Soc.,} in press.}

\def\pasp##1,{{\it Pub.~A.S.P.,\ }{\bf##1},}
\def\pasppr{{\it Pub.~A.S.P.,} in press.}
\def\ajpasp##1,{{\it Publ. Astron. Soc. Pac.,\ }{\bf##1},}
\def\ajpasppr{{\it Publ. Astron. Soc. Pac.,} in press.}
}

\def\twelvepoint{
  \font\twelverm=cmr12
  \font\twelvei=cmmi12
  \font\twelvesy=cmsy10 scaled\magstep1
  \font\tenrm=cmr10
  \font\teni=cmmi10
  \font\tensy=cmsy10
  \font\sevenrm=cmr7
  \font\seveni=cmmi7
  \font\sevensy=cmsy7
  \font\it=cmti12
  \font\bf=cmbx12 
  \font\sl=cmsl12
  \textfont0= \twelverm \scriptfont0=\tenrm 
\scriptscriptfont0=\sevenrm
  \def\rm{\fam0 \twelverm}   
  \textfont1=\twelvei  \scriptfont1=\teni 
\scriptscriptfont1=\seveni
  \def\mit{\fam1 } \def\oldstyle{\fam1 \twelvei}
  \textfont2=\twelvesy \scriptfont2=\tensy 
\scriptscriptfont2=\sevensy
\def\doublespace{\baselineskip=24pt\lineskip=0pt 
\lineskiplimit=-5pt}
\def\singlespace{\baselineskip=13.5pt\lineskip=0pt
\lineskiplimit=-5pt}
\def\oneandahalf{\baselineskip=18pt\lineskip=0pt
\lineskiplimit=-5pt}
}

\def\eighteenpoint{
  \font\eighteenrm=cmr10 scaled\magstep3
  \font\eighteeni=cmmi10 scaled\magstep3
  \font\eighteensy=cmsy10 scaled\magstep3
  \font\eighteenrm=cmr10 scaled\magstep3
  \font\twelvei=cmmi12  
  \font\twelvesy=cmsy12
  \font\teni=cmmi10
  \font\tensy=cmsy10  
  \font\seveni=cmmi7
  \font\sevensy=cmsy7
  \font\it=cmti10 scaled \magstep3
  \font\bf=cmb10 scaled \magstep3
  \font\sl=cmsl10 scaled \magstep3
  \textfont0= \eighteenrm \scriptfont0=\twelverm 
\scriptscriptfont0=\tenrm
  \def\rm{\fam0 \eighteenrm}   
  \textfont1=\eighteeni  \scriptfont1=\twelvei 
\scriptscriptfont1=\teni
  \def\mit{\fam1 } \def\oldstyle{\fam1 \eighteeni}
  \textfont2=\eighteensy \scriptfont2=\tensy 
\scriptscriptfont2=\sevensy
\def\doublespace{\baselineskip=30pt\lineskip=0pt
\lineskiplimit=-5pt}
\def\singlespace{\baselineskip=20pt\lineskip=0pt
\lineskiplimit=-5pt}
\def\oneandahalf{\baselineskip=25pt\lineskip=0pt
\lineskiplimit=-5pt}
\def\deg{^{\raise2pt\hbox{$\circ$}}}}


\def\nH2{$n_{\rm H_2}~$}
\def\nh2{$n_{\rm H_2}~$}
\def\H2{${\rm H_2}~$}
\def\h2{${\rm H_2}~$}
\def\cm{{\rm cm}}
\def\cm3{cm$^{-3}$}
 
\def\Tbr{$T_{br}~$}
\def\T{$T_{\rm H_2}~$}
\def\pil{$^2\Pi_{3/2}$}
\def\pir{$^2\Pi_{1/2}$}
\def\fo{$f_{\rm ortho-H_2}$}

\def\fOH{$f_{\rm OH}$}

\begintex
\centerline{\bf 5 cm OH MASERS AS DIAGNOSTICS OF PHYSICAL CONDITIONS}
\centerline{\bf IN STAR-FORMING REGIONS. }
\bigskip
\centerline{KONSTANTINOS G. PAVLAKIS$^{1,2,3}$ \& NIKOLAOS D. KYLAFIS$^{2,3}$}
\bigskip
\leftline {$^1$University of Leeds, Department of Physics and Astronomy,
  Woodhouse Lane, Leeds LS2 9JT}
\leftline {$^2$University of Crete, Physics Department, 714 09 Heraklion, 
Crete, Greece}
\leftline{$^3$Foundation for Research and Technology-Hellas, P.O. Box 1527, 
711 10 Heraklion, Crete, Greece}

\centerline{pavlakis@ast.leeds.ac.uk,  kylafis@physics.uch.gr}
\bigskip

\centerline{{\it Received:}~~\underbar{~~ \ \ \ \ \ \ \ ~~}{\it~;}
{\it accepted:}~~\underbar{~~ \ \ \ \ \ \ \ ~~}}
\bigskip

\centerline{ABSTRACT}
\bigskip

We demonstrate that the observed characteristics of the
5 cm OH masers in star-forming regions can be 
explained with the same model and the same parameters as the 18 cm and the 
6 cm OH masers.  In our already published study of the 18 cm and the 6 cm
OH masers in star-forming regions we had examined the pumping of the 5 cm
masers, but did not report the results we had found because of some missing
collision rate coefficients, which in principle could be important.  The
recently published observations on the 5 cm masers of OH encourage us to
report our old calculations along with some new ones that we have performed.
These calculations, in agreement with the observations, reveal the
main lines at 5 cm as strong masers, the 6049 MHz satellite line as a weak
maser, and the 6017 MHz satellite line as never inverted for reasonable
values of the parameters.

\bigskip \noindent
{\it Subject headings:} ISM: molecules --- masers --- molecular processes 
--- radiative transfer --- stars: formation
\vfill \eject
\centerline{1. INTRODUCTION}
\bigskip

The OH molecules in star-forming regions are rich in maser emission. 
They exhibit: 
a) Four maser lines (at 18 cm) in the ground state \pil, $J=3/2$ 
(e.g., Gaume \& Mutel 1987; Cohen, Baart, \&
Jonas 1988; for reviews see Reid \& Moran 1981; Cohen 1989; Elitzur 1992).
b) Three maser lines (at 5 cm) in the first excited state \pil, $J=5/2$ 
(Knowles, Caswell, \& Goss 1976; Guilloteau et al. 1984; 
Smits 1994; Caswell \& Vaile 1995;
Baudry et al. 1997; Desmurs et al. 1998; Desmurs \& Baudry 1998).
c) Three maser lines (at 6 cm) in the next level \pir, $J=1/2$ 
(Gardner \& Martin Pintado 1983; 
Gardner \& Whiteoak 1983; Palmer, Gardner, \& Whiteoak 1984;
Gardner, Whiteoak, \& Palmer 1987; Baudry et al. 1988; Baudry \& Diamond 1991;
Cohen, Masheder, \& Walker 1991).
d) One maser line (at 2 cm) in the level \pil, $J=7/2$ 
(Turner, Palmer, \& Zuckerman 1970; Baudry et al. 1981; 
Baudry \& Diamond 1998).

In our recently reported calculations (Pavlakis \& Kylafis 1996a, hereafter
Paper I; Pavlakis \& Kylafis 1996b, hereafter Paper II) we explained
theoretically the observed characteristics of the 18 cm and the 6 cm maser
lines of OH.  Naturally, we had computed also the maser emission of the 5 cm 
lines of OH, but we decided to ``not show or discuss the maser lines in the
excited state \pil, $J=5/2$ because this state is directly connected with the
state \pir, $J=7/2$, which is not included in our calculations.
We urge quantum chemists to compute collision rate coefficients for as 
many transitions as possible'' (Paper I).  

Soon after our calculations were published, Baudry et al. (1997) reported 
an extensive study of the 5 cm maser lines of OH in star-forming regions.
To our surprise, {\it our already performed calculations explain the
observed characteristics.}  This probably means that, for temperatures
between 100 and 200 K that are thought appropriate for OH maser regions,
the missing collision rate coefficients are not important for the 5 cm masers.
Thus, we are encouraged to publish 
our results.  Of course, when the missing collision rate coefficients are 
computed, it will be reassuring to show that they are indeed not important 
for the 5 cm masers.

In \S 2 we discuss briefly the model that we used,
in \S 3 we present the results of the calculations, 
in \S 4 we compare our calculations with the observations and
in \S 5 we present our conclusions.

\bigskip
\centerline{2. MODEL}
\bigskip

Our model is the same as that in Papers I and II.  Not only this,
but {\it the values of the parameters are exactly the same.}  
Thus, no parameters are adjusted for any qualitative or quantitative 
explanation of the observations.

The maser regions are modeled as cylinders of length $l=5 \times 10^{15}$ cm
and diameter $d=10^{15}$ cm.  The characteristic bulk velocity
in the maser region is denoted by $V$ and the assumed velocity field there
is given 
by $ \vec {\bf v} =V/(d/2) \rho {\bf\hat \rho} +(V/l) 
z {\bf\hat z}$, where ($\rho, z$) are the cylindrical coordinates,
and ${\bf\hat\rho}$ and ${\bf\hat z}$ are the corresponding unit vectors.

The fractional abundances of OH and ortho-H$_2$ with
respect to density of H$_2$ molecules in the maser region
are denoted by \fOH \ and \fo, respectively, while the density 
of H$_2$ molecules and the 
kinetic temperature there are denoted by \nH2 and $T_{\rm H_2}$, respectively.
Finally, the brightness temperature of the maser lines is denoted by
$T_{br}$, the dilution factor of the far infrared radiation field by $W$
(see eq. [5] of Paper II), 
the dust optical depth parameter by $p$
(see eq. [4] of Paper II), 
and the dust temperature by $T_d$.

In addition to the exploration of the parameters used in Papers I and II,
we also explore here the effects of the fractional abundance of OH.

In the figures of this paper the key is as follows:
The brightness temperature of the 6049 MHz transition is shown as a solid 
line, that of the 6035 MHz transition by a dotted line, that of the 6017 MHz
by a dashed line and that of the 6031 MHz by a dot-dashed line.

\bigskip
\centerline{3. CALCULATIONS AND PRESENTATION OF RESULTS}
\bigskip
\centerline{3.1.  Collisions Only}
\bigskip

For kinetic temperatures $100 \ltorder T_{\rm H_2} \ltorder 200$ K, which are 
thought to be prevailing in H II/OH maser regions, there are several locally 
(i.e., thermally) overlapping lines of OH.  Nevertheless, it is interesting 
to look at calculations, which take into account collisions only, in order to 
see what their effects are on the pumping of OH molecules (see also Paper I).
Interestingly, we have found that for temperatures $T_{\rm H_2} \ltorder 150$ 
K, which are highly likely for H II/OH regions, the effects of locally 
overlapping lines are insignificant on the pumping of the 5 cm maser lines
and their inclusion changes the results 
of our calculations by less than a factor of two.  Thus, if there are no 
large velocity gradients in the maser regions and the external FIR
field is weak, collisions alone determine the 5 cm maser emission
of OH at $T_{\rm H_2} \ltorder 150$ K.

We have found that collisions alone are unable to invert the main lines
at 5 cm.  For \fo$=1$ and \fOH$=10^{-5}$ 
only the 6049 MHz satellite line is masing for  hydrogen
densities $2 \times 10^5 \ltorder n_{\rm H_2} \ltorder 7 \times 10^6$ 
cm$^{-3}$. 
The peak of the brightness temperature occurs at $n_{\rm H_2} \sim 10^6$ 
cm$^{-3}$ and it is $T_{br} \sim 10^9$ K above $T_{\rm H_2} = 100$ K
(see Figure 1 below and the discussion in the next subsection).

As \fo decreases, the brightness temperature of the 6049 MHz line decreases
faster than exponential. For \fo$=0.5$ the peak of the 
brightness temperature is
$T_{br} = 6 \times 10^6$ K and it is at the limits of dedectability.
For values of \fo below 0.2 the inversion disappears and no 5 cm line
shows inversion.

\bigskip
\centerline{3.2. Collisions and Local Line Overlap}
\bigskip

The effects of collisions and local line overlap 
{\it cannot} be separated. It is simply a good fortune that for temperatures 
up to about 150 K 
the effects of
collisions dominate 
those of
local line overlap.

Assuming that large velocity gradients and a significant FIR radiation 
field are absent in the maser regions (see below for their effects), 
we have computed the 5 cm OH
maser emission as a function of \nH2 taking into account both collisions and
local line overlap.
For \T$= 150$ K, \fo$=1$, $f_{\rm OH} = 10^{-5}$ and $V=0.6$ km s$^{-1}$
(for which we do not have any non-locally overlapping lines),
we show in Figure 1 the brightness temperature \Tbr of the 6049 MHz line
(the only inverted OH line at 5 cm) as a function of \nH2. 
This is a 
quite strong maser line with a peak brightness temperature $T_{br} = 6 
\times 10^8$ K.
As the temperature \T
increases further, more and more pairs of lines overlap locally and their
degree of overlap also increases. For temperatures up to 170 K, local
overlap causes only quantitative (not qualitative) changes in the results.
The peak brightness temperature decreases with increasing kinetic temperature 
and the range of densities over which inversion occurs also
decreases. For \T$=170$ K, the peak $T_{br}$ of the 6049 MHz maser line
falls to $10^7$ K 
and inversion occurs for $10^5 \ltorder n_{\rm H_2} \ltorder 10^6$ cm$^{-3}$.

Above \T$=170$ K, the effects of local line overlap introduce qualitative
changes.  
The 6049 MHz line continues to weaken, while the 6035 MHz line now
appears.
As the temperature approaches 200 K, there are fifteen pairs and one
triple of locally overlapping lines. Figure 2 shows \Tbr as a function of
\nH2 for \T$=200$ K, \fo$=1$, $f_{\rm OH} = 10^{-5}$ and $V=0.6$ km s$^{-1}$.
At this relatively high temperature, the peak \Tbr of the 6049 MHz maser
line is only $10^6$ K, 
while for the 6035 MHz main line, 
\Tbr $\sim 10^9$ K at $n_{\rm H_2} \sim 10^7$ cm$^{-3}$.

When \fo$=0$, no OH 5 cm maser line appears
for kinetic temperatures lower than 170 K.
For \T$=200$ K, \fo$=0$, \fOH$=10^{-5}$ and $V=0.6$ km s$^{-1}$
the results are shown in Figure 3. The peak \Tbr of the 6035 MHz
line is one order of magnitude stronger than that for \fo$=1$, but the 
6049 MHz line is absent.  Thus, as with the 1720 MHz maser line (see Paper I),
the 6049 MHz maser line could be a diagnostic (but see below) of the
abundance of ortho-H$_2$ in maser regions.  

The fractional 
abundance of OH in star-forming regions is probably not constant
independent of density.  To explore this possibility we have computed models
with \fOH$=10^{-6}$ 
(the results are not shown in a Figure).
No qualitative changes occur in comparison with
the results for \fOH$=10^{-5}$ (see Figures 2 and 3).  The 6035 MHz line
is of the same intensity as in Figures 2 and 3, but it is inverted at 
densities a factor of three higher.  The 6049 MHz line is reduced in 
intensity to the point of being unobservable ($T_{br} \ltorder 10^3$ K).  
For this line also the inversion occurs at densities a factor of three
higher than those in Figure 2.

\bigskip
\centerline{3.3.  Collisions, Local and Non-local Line Overlap}
\bigskip

In addition to collisions and local line overlap, we now take into account
the effects of non-local line overlap (for details see Paper II).
In order to avoid showing a multitude of models, we take the representative 
value of 150 K for $T_{\rm H_2}$.  For \fo $ = 1$ and \fOH$= 10^{-5}$,
we start with a characteristic velocity $V=1$ km s$^{-1}$, for 
which the effects of non-local line overlap are already important. Figure 4 
shows the brightness temperature of the masing lines as a function of \nH2.
Comparing Figure 4 with Figure 1 we see the dramatic effects that non-local
line overlap has on the maser transitions. 
For relatively high hydrogen densities, two 5 cm lines
are inverted. The 6035 MHz main line with high brightness temperature 
and the 6017 MHz satellite line.

Increasing the characteristic velocity to $V=2$ km s$^{-1}$, but keeping the
rest of the parameters the same, results in a significant reduction of the 
6049 MHz line (Figure 5). For \nH2$\sim 10^8$ \cm3, the 6031 MHz main
line and the 6017 MHz satellite one are inverted with high 
brightness temperatures. At even higher densities the 6035 MHz main line
is inverted, the 6017 MHz line remains strongly inverted, while the 6031 MHz
one is suppressed.   A further increase of the velocity to
$V=3$ km s$^{-1}$ results in the complete disappearance of the 6049 MHz
line (Figure 6).  

As in the previous subsection, a significant 
reduction of the abundance of ortho-H$_2$ 
results in the disappearance of the 6049 MHz line as a maser line.  This is
true for $V=1$, 2 or 3 km s$^{-1}$.  As a characteristic example we show
the case of \fo$=0$, \fOH$= 10^{-5}$ and $V=2$ km s$^{-1}$ (Figure 7). 
Below \nH2$\sim 3 \times 10^7$ \cm3 no maser line appears.

Finally, a reduction of \fOH~ by an order of magnitude has the general result 
of significantly reducing the brightness temperature of the 6049 MHz line
($T_{br} \ltorder 10^4$ K),
as it was also seen in the previous subsection. Furthermore, our calculations
have shown that \fOH$= 10^{-6}$ results in destroying the inversion of all
5 cm maser lines at relatively high densities 
(\nH2$\gtorder 4 \times 10^7$ \cm3).
 
\bigskip
\centerline{3.4. Effects of a FIR Radiation Field}
\bigskip

From the calculations presented so far, it is evident that the 5 cm {\it main 
lines} of OH {\it are never inverted together} for densities thought 
prevailing in star-forming regions (i.e., \nH2$\ltorder {\rm few} \times 
10^7$ \cm3),
when a far infrared (FIR) radiation field is absent.  In this subsection we 
will demonstrate that a FIR radiation field is necessary to reproduce the 
observed features of the 5 cm lines and their correlations with the ground
state $^2\Pi_{3/2}$, $J=3/2$ and the excited state $^2\Pi_{1/2}$, $J=1/2$ 
OH masers.

For \T $ =150$ K, $f_{\rm OH}= 10^{-5}$, $f_{\rm ortho-H_2}$=1, $V$=1 
km s$^{-1}$ and dilution factor $W=0.01$ (see Paper II), 
the main lines at 5 cm are inverted at low densities 
(\nH2$\ltorder {\rm 
few} \times 10^7$ \cm3)
when
$T_d > T_{\rm H_2}$.  When $T_d < T_{\rm H_2}$ (see Figures 8a and 8b) the 
results are similar  i.e., differences less than a factor of 2 in the
$T_{br}$ to those of Figure 4, where there was no external 
FIR radiation field.  However, when $T_d > T_{\rm H_2}$ 
(see Figures 8c and 8d), 
the main line at 6035 and 6031 MHz make their appearance as masers.

Increasing the strength of the FIR radiation field by taking $W=0.1$ has
dramatic effects on the 5 cm lines of OH.
Figures 9a - 9d show $T_{br}$ of the maser lines as a function of 
$n_{\rm H_2}$ for \T $ =150$ K, $f_{\rm OH}= 10^{-5}$,
$f_{\rm ortho-H_2}$=1, $V$=1 km s$^{-1}$ and dilution factor
W=0.1.  Both 5 cm main lines are masing with the 6035 MHz line stronger than
the 6031 MHz one in the range of densities where both lines are inverted.
The satellite line at 6049 MHz is also masing but the other satellite line
at 6017 MHz is never inverted for \nH2$\ltorder {\rm few} \times 10^7$ 
\cm3.

Remarkably, the abundance of ortho-H$_2$ causes no 
changes as to which 5 cm lines of OH are masers. 
Figures 10a - 10d are made with \fo$=0$ and the rest
of the parameters the same as in Figures 9a - 9d, respectively.

What has dramatic effects on the pumping of the 5 cm lines is the 
characteristic velocity.  For $V=2$ km s$^{-1}$ and $V=3$ km s$^{-1}$
with the rest of the parameters the same as in Figures 9a - 9d, the results
are shown in Figures 11a - 11d and 12a - 12d, respectively.
As it is clear from these Figures, an increase of $V$ (i.e., 
increase of nonlocal overlap) causes suppression of the 5 cm main lines.
The 6035 MHz line is not inverted at all. The 6031 MHz either is not
inverted (compare Figs. 9a and 12a) or it is much weaker (compare 
Figs. 9d and 12d) depending on the strength of the FIR field. The reader
should notice the competition between the FIR field, which inverts the main 
lines (as we go from a to d in Figs. 9, 11, and 12 the FIR field increases)
and the nonlocal overlap, which suppresses the inversion (as we go from 
Fig. 9 to Fig. 11 and then to Fig. 12 the nonlocal overlap increases).

For completeness, we take one of our 
cases that agrees qualitatively well with the observational data, 
namely
the case presented in Figure 9c, and explore the effects of the abundance
of OH.  For \fOH$=10^{-4}$ and \fOH$=10^{-6}$ the results are shown in
Figures 13 and 14, respectively.  The abundance of OH introduces only 
quantitative changes.  An enhanced abundance of OH increases the brightness
temperature of the 6049 MHz maser line, while a reduced abundance has the 
opposite effect.  The main lines at 6035 and 6031 MHz remain essentially
unaffected.

Since the effects of the abundance of OH on the ground state 
$^2\Pi_{3/2}$, $J=3/2$ and the excited state $^2\Pi_{1/2}$, $J=1/2$ were not
investigated in Paper II, we show in Figure 15 the case of \fOH$=10^{-6}$ and 
all the other parameters the same as in Figure 6c of Paper II.
 
\bigskip
\centerline{4. COMPARISON WITH OBSERVATIONS}  

Since the original discovery of emission from the \pil, $J=5/2$
(Yen et al. 1969), many surveys were made for detection of 5 cm maser
emission toward a variety of sources (Knowles et al. 1976; Guilloteau et al. 
1984; Smits 1994). Caswell and Vaile (1995) surveyed for 6035 MHz masers in
208 OH sources with peak 1665MHz flux density greater than 0.8 Jy.  
Only 35 masers at 6035 MHz were detected, `` a result that agrees well 
with our calculations ''. Since these observations were made with
a single dish, and the authors have not proven that any of the 1665-6035
MHz ``pairs'' come from the same region, these observations must be 
interpreted solely as a tendency of the 1665 MHz line to be inverted
more easily than the 6035 MHz one. Our results qualitatively agree with
this. 
The 1665 MHz line (see Paper II) is inverted in a much
broader range of densities, velocity fields and strengths of a FIR field
than the 6035MHz line. 

Let's for the rest of this section 
restrict to our results in the
presence of a FIR field, $V < 1.5$ km s$^{-1}$ and $n_{\rm H_2} < 10^7$ 
cm$^{-3}$. The 6035 MHz line is weaker than the 1665 MHz one and is inverted
in a range of densities which is a subset 
of the range of densities over which the 1665 MHz line is 
inverted. As the FIR field gets stronger, this subregion becomes broader 
and the 6035 MHz line tends to be inverted in the same range of densities
as the 1665 MHz line.  By taking also into account our result that the 
stronger the FIR radiation field is the stronger the 1665 and 6035 MHz masers 
are, and assuming that both lines come from the same region, 
our models are in qualitative agreement with the observational result of
Caswell and Vaile (1995) that the greater the peak of 1665 MHz maser intensity,
the greater the detection rate of 6035 MHz masers.

An extensive search for all four maser lines in 5 cm has been made 
by Baudry et al. (1997)
toward 265 strong FIR sources and the general observed characteristics
of these 5 cm masers can be explained by our calculations. Their observations
show (see also Desmurs et al. 1998) that  
the main-line masers at 6035 MHz, in the $^2\Pi_{3/2}$, 
$J=5/2$ state of OH, are generally stronger and more common than 
those at 6031 MHz in H II/OH regions. 
Nevertheless the 6031 MHz line is frequently observed to be masing.
Strong 5 cm satellite line masers are not observed in the $J=5/2$ state 
of OH. The 6017 MHz line is often found in absorption while the other
satellite line at 6049 MHz is observed in weak emission which could 
correspond to low gain masers.   Our theoretical calculations are in 
good qualitative agreement with these observations.  
The 6017 MHz line is never 
inverted in our calculations and the 6049 MHz line is weak in a wide range
of parameters thought to be prevailing in star-forming regions.

Our calculations show that a combination of a FIR radiation field, collisions 
and line overlap is necessary to reproduce the general features of 6 GHz 
H II/OH masers. Nevertheless, simultaneous or nearly simultaneous
VLBI observations at 1.6 GHz,
4.7 GHz and 6 GHz are necessary to restrict the range of parameters for
the inversion of these masers and a search of the correlation between 
5 cm maser and FIR radiation field strength would be important to prove 
or not the importance of FIR radiation for the inversion of these masers.  

\bigskip
\centerline{5. SUMMARY AND CONCLUSIONS}
\bigskip
We have performed a detailed, systematic study of OH maser pumping in order
to attempt to invert the problem and from the OH maser observations to infer
the physical conditions in H II/OH regions.  This was partially accomplished
in Papers I and II.  With the present study of the 5 cm masers of OH 
the predictions of our model are:

1) When strong 5 cm maser main lines are seen, a FIR radiation field must
be strong there, 
i.e., high value of $W$ or $p$ or $T_d$ or a combination of them.

2) Inversion of both main lines at 5 cm requires relatively small 
velocity gradients. For $V \ltorder 1$ km s$^{-1}$
and a FIR radiation field present, these lines are always seen. If these lines 
are seen together in the same spatial region, the 1665 MHz OH ground state
main line maser will also be observed in the same region, while there is a
high probability the other ground state main line maser at 1667 MHz to be 
observed too (see Paper II).

3) When the 6031 MHz maser line is observed in a region where there is no 
detection of 6035 MHz maser, the 1665 MHz ground state line is
inverted in the same spatial region. This situation has a great probability
to be indicative of relatively large velocity gradients ($V > 1$ km s$^{-1}$). 

4) When the 6049 MHz maser line is seen as a strong line (say, as strong as the
5 cm main lines are typically seen), then \fOH$\gtorder 10^{-5}$.

5) We predict that maser spots showing very strong 18 cm main lines should
exhibit 5 cm maser main lines also.  This may have already been seen 
(Caswell \& Vaile 1995), but VLBI observations are needed to confirm or
reject our prediction. 

6) We also predict that 18 cm maser main lines with $V \gtorder 2$ km s$^{-1}$
will not be accompanied by 5 cm maser main lines. 

This research has been supported in part by a grant from the General 
Secretariat of Research and Technology of Greece and a Training and Mobility
of Researchers Fellowship of the 
European Union under contract No ERBFMBICT972277.

\vfill \eject
\centerline{REFERENCES}

\beginrefs

Baudry, A., \& Diamond, P. J. 1991, A\&A, 247, 551

------------ 1998, A\&A, 331, 697

Baudry, A., Desmurs, J. F., Wilson, T. L., \& Cohen, R. J. 1997, A\&A, 
325, 255

Baudry, A., Diamond, P. J., Booth, R. S., Graham, D., \& Walmsley, C. M.
1988, A\&A, 201, 105

Baudry, A., Walmsley, C. M., Winnberg, A., \& Wilson, T. L. 1981, A\&A, 
102, 287

Caswell, J. L., \& Vaile, R. A. 1995, MNRAS, 273, 328

Cohen, R. J. 1989, Rep. Prog. Phys., 52, 881

Cohen, R. J., Baart, E. E., \& Jonas, J. L. 1988, MNRAS, 231, 205

Cohen, R. J., Masheder, M., \& Walker, R. N. F. 1991, MNRAS, 250, 611

Desmurs, J. F., Baudry, A., Wilson, T. L., Cohen, R. J., \& Tofani, G.
1998, A\&A, 334, 1085

Desmurs, J. F., \& Baudry, A. 1998, A\&A, in press

Elitzur, M. 1992, ARA\&A, 30, 75

Gardner, F. F. \& Martin-Pintado, J. 1983, A\&A, 121, 265

Gardner, F. F., \& Whiteoak, J. B. 1983, MNRAS, 205, 297

Gardner, F. F., Whiteoak, J. B., \& Palmer, P. 1987, MNRAS, 225, 469

Gaume, R. A., \& Mutel, R. L. 1987, ApJ Suppl., 65, 193

Guilloteau, S., Baudry, A., Walmsley, C. M.,  Wilson, T. L., \& Winnberg, A.
1984, A\&A, 131, 45

Knowles, S. H., Caswell, J. L., \& Goss, W. M. 1976, MNRAS, 175, 537
 
Palmer, P., Gardner, F. F., \& Whiteoak, J. B. 1984, MNRAS, 211, 41p

Pavlakis, K. G., \& Kylafis, N. D. 1996a, ApJ, 467, 300 (Paper I)

------------ 1996b, ApJ, 467, 309 (Paper II)

Reid, M. J., \& Moran, J. M. 1981, ARA\&A, 19, 231

Smits, D. P. 1994, MNRAS, 269, L11

Turner, B. E., Palmer, P., \& Zuckerman, B. 1970, ApJ, 160, L125

Yen, J. L., Zuckerman, B., Palmer, P., \& Penfield, H. 1969, 
ApJ 156, L27

\endrefs

\vfill \eject
\centerline{FIGURE CAPTIONS}
\noindent
FIG. 1.--- Brightness temperature \Tbr as a function of density \nH2 for
\T$=150$ K, \fo $ =1$, \fOH$=10^{-5}$ and $V = 0.6$ km s$^{-1}$.
The values of the other parameters are given in \S 2.
\bigskip \noindent
FIG. 2.--- Same as in Figure 1, but \T$=200$ K.
\bigskip \noindent
FIG. 3.--- Same as in Figure 2, but \fo$=0$.
\bigskip \noindent
FIG. 4.--- Brightness temperature \Tbr as a function of density \nH2 for
\T$=150$ K, \fo $ =1$, \fOH$=10^{-5}$ and $V = 1$ km s$^{-1}$.
The values of the other parameters are given in \S 2.
\bigskip \noindent
FIG. 5.--- Same as in Figure 4, but for $V = 2$ km s$^{-1}$.
\bigskip \noindent
FIG. 6.--- Same as in Figure 4, but for $V = 3$ km s$^{-1}$.
\bigskip \noindent
FIG. 7.--- Same as in Figure 5, but for \fo$=0$.
\bigskip \noindent
FIG. 8a.--- Brightness temperature \Tbr as a function of density \nH2 for
\T$=150$ K, \fo$=1$, \fOH$=10^{-5}$, $V=1$ km s$^{-1}$, $T_d=100$ K, $p=1$
and $W=0.01$.  The values of the other parameters are given in \S 2.
\bigskip \noindent
FIG. 8b.--- Same as in Figure 8a, but for $p=2$.
\bigskip \noindent
FIG. 8c.--- Same as in Figure 8a, but for $T_d = 200$ K.
\bigskip \noindent
FIG. 8d.--- Same as in Figure 8b, but for $T_d = 200$ K.
\bigskip \noindent
FIG. 9a.--- Same as in Figure 8a, but for $W=0.1$.
\bigskip \noindent
FIG. 9b.--- Same as in Figure 8b, but for $W=0.1$.
\bigskip \noindent
FIG. 9c.--- Same as in Figure 8c, but for $W=0.1$.
\bigskip \noindent
FIG. 9d.--- Same as in Figure 8d, but for $W=0.1$.
\bigskip \noindent
FIG. 10a.--- Same as in Figure 9a, but for \fo$=0$.
\bigskip \noindent
FIG. 10b.--- Same as in Figure 9b, but for \fo$=0$.
\bigskip \noindent
FIG. 10c.--- Same as in Figure 9c, but for \fo$=0$.
\bigskip \noindent
FIG. 10d.--- Same as in Figure 9d, but for \fo$=0$.
\bigskip \noindent
FIG. 11a.--- Same as in Figure 9a, but for $V=2$ km s$^{-1}$.
\bigskip \noindent
FIG. 11b.--- Same as in Figure 9b, but for $V=2$ km s$^{-1}$.
\bigskip \noindent
FIG. 11c.--- Same as in Figure 9c, but for $V=2$ km s$^{-1}$.
\bigskip \noindent
FIG. 11d.--- Same as in Figure 9d, but for $V=2$ km s$^{-1}$.
\bigskip \noindent
FIG. 12a.--- Same as in Figure 9a, but for $V=3$ km s$^{-1}$.
\bigskip \noindent
FIG. 12b.--- Same as in Figure 9b, but for $V=3$ km s$^{-1}$.
\bigskip \noindent
FIG. 12c.--- Same as in Figure 9c, but for $V=3$ km s$^{-1}$.
\bigskip \noindent
FIG. 12d.--- Same as in Figure 9d, but for $V=3$ km s$^{-1}$.
\bigskip \noindent
FIG. 13.--- Same as Figure 9c, but for \fOH$=10^{-4}$.
\bigskip \noindent
FIG. 14.--- Same as Figure 9c, but for \fOH$=10^{-6}$.
\bigskip \noindent
FIG. 15.--- Same as Figure 6c of Paper II, but for \fOH$=10^{-6}$.

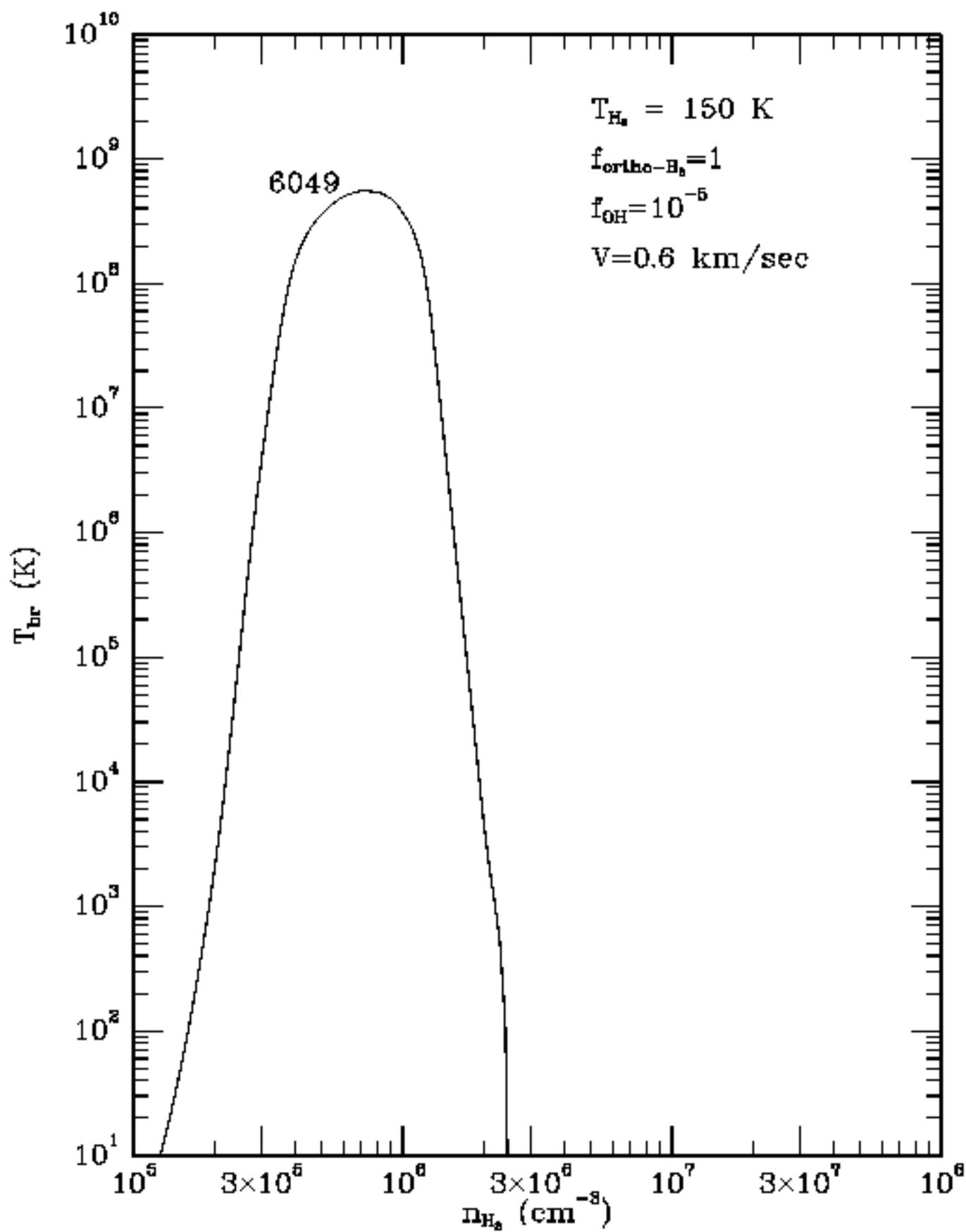
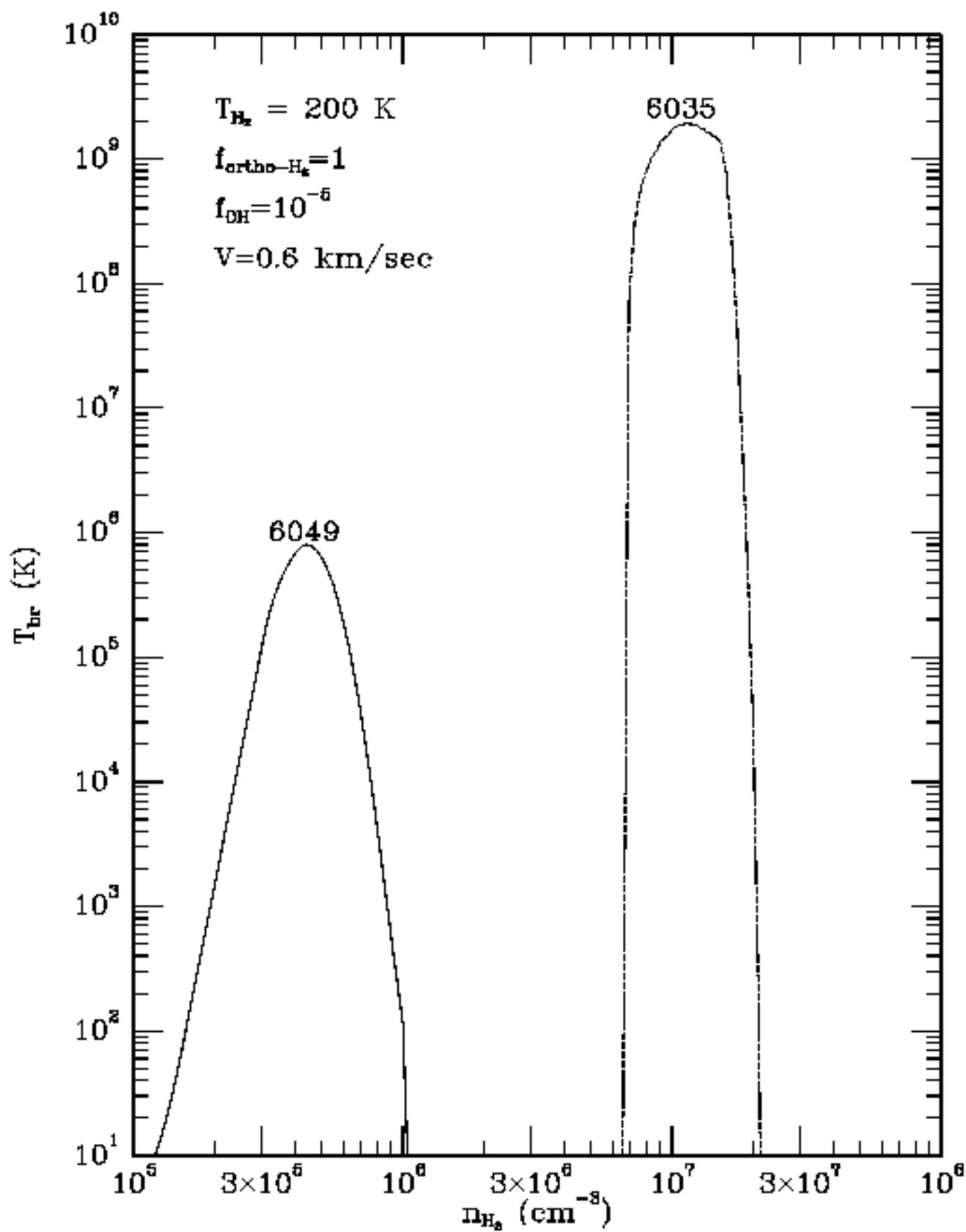
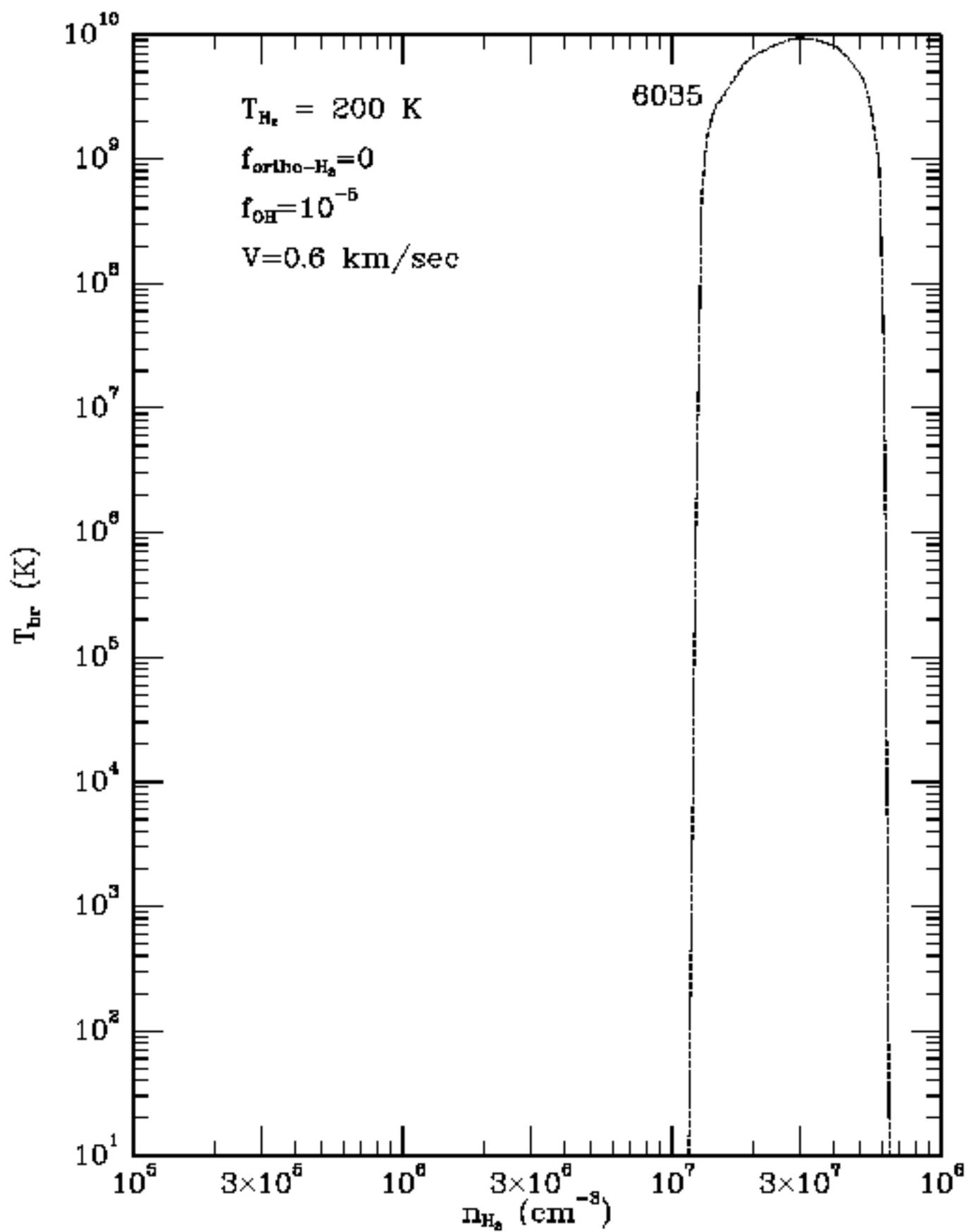
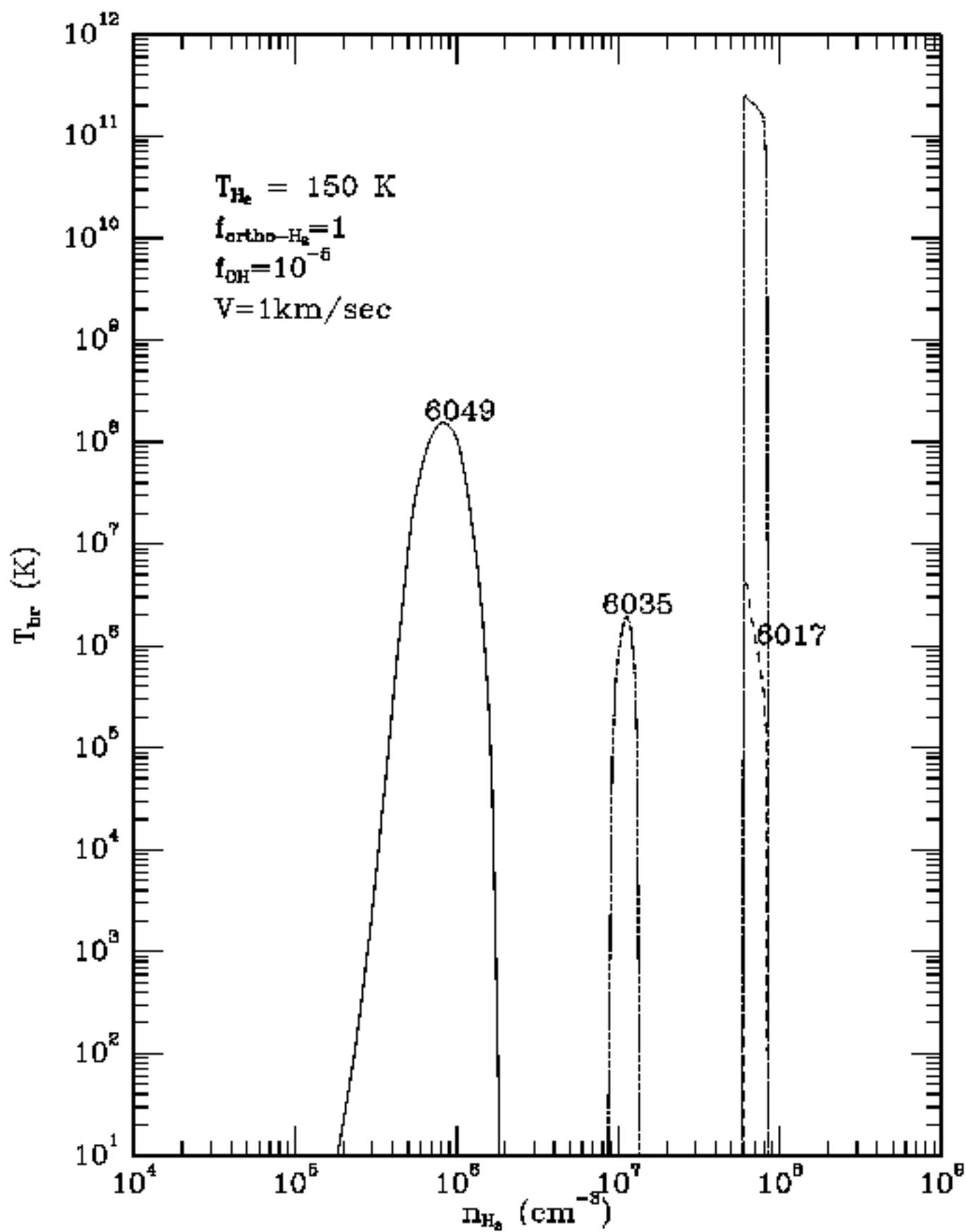
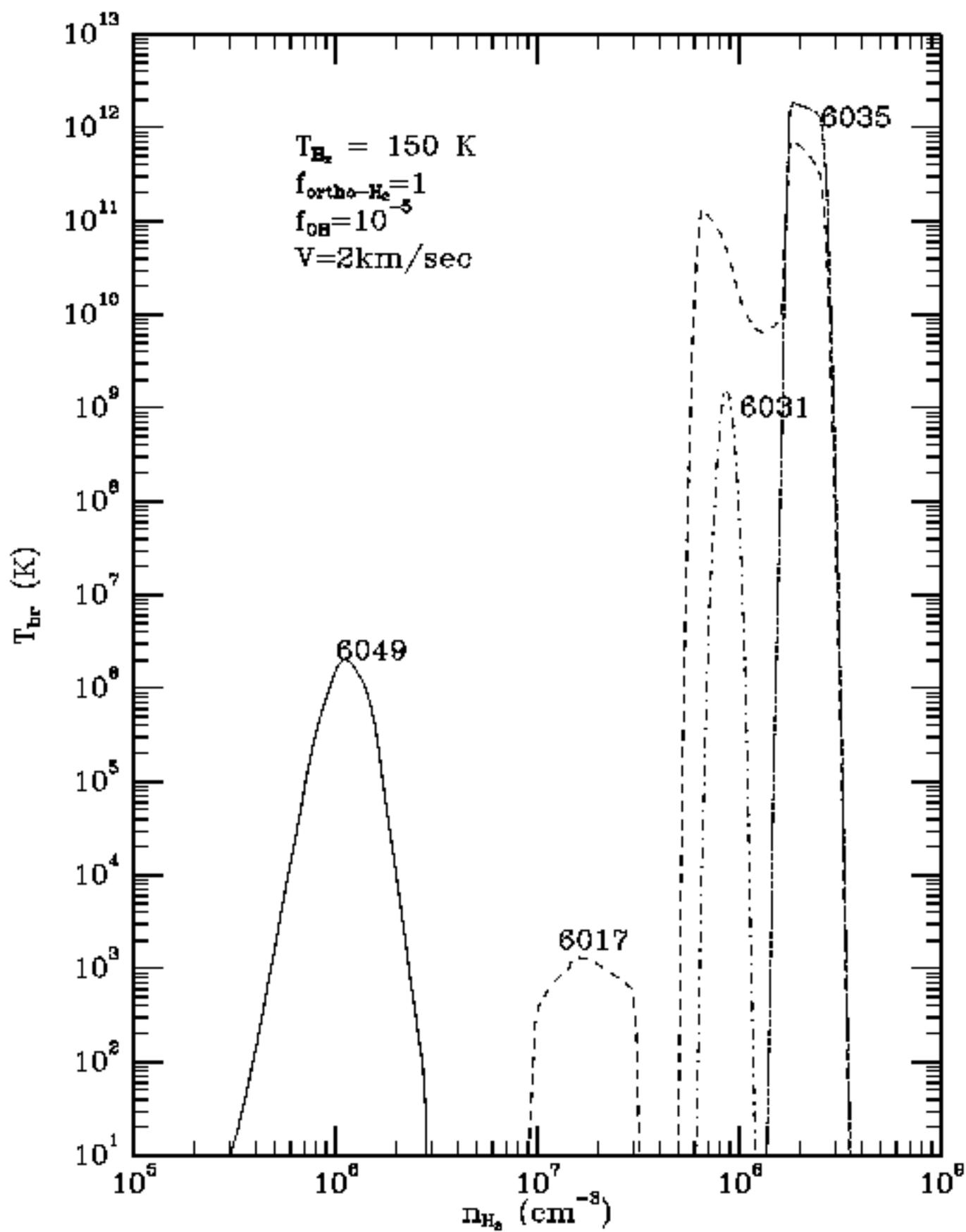
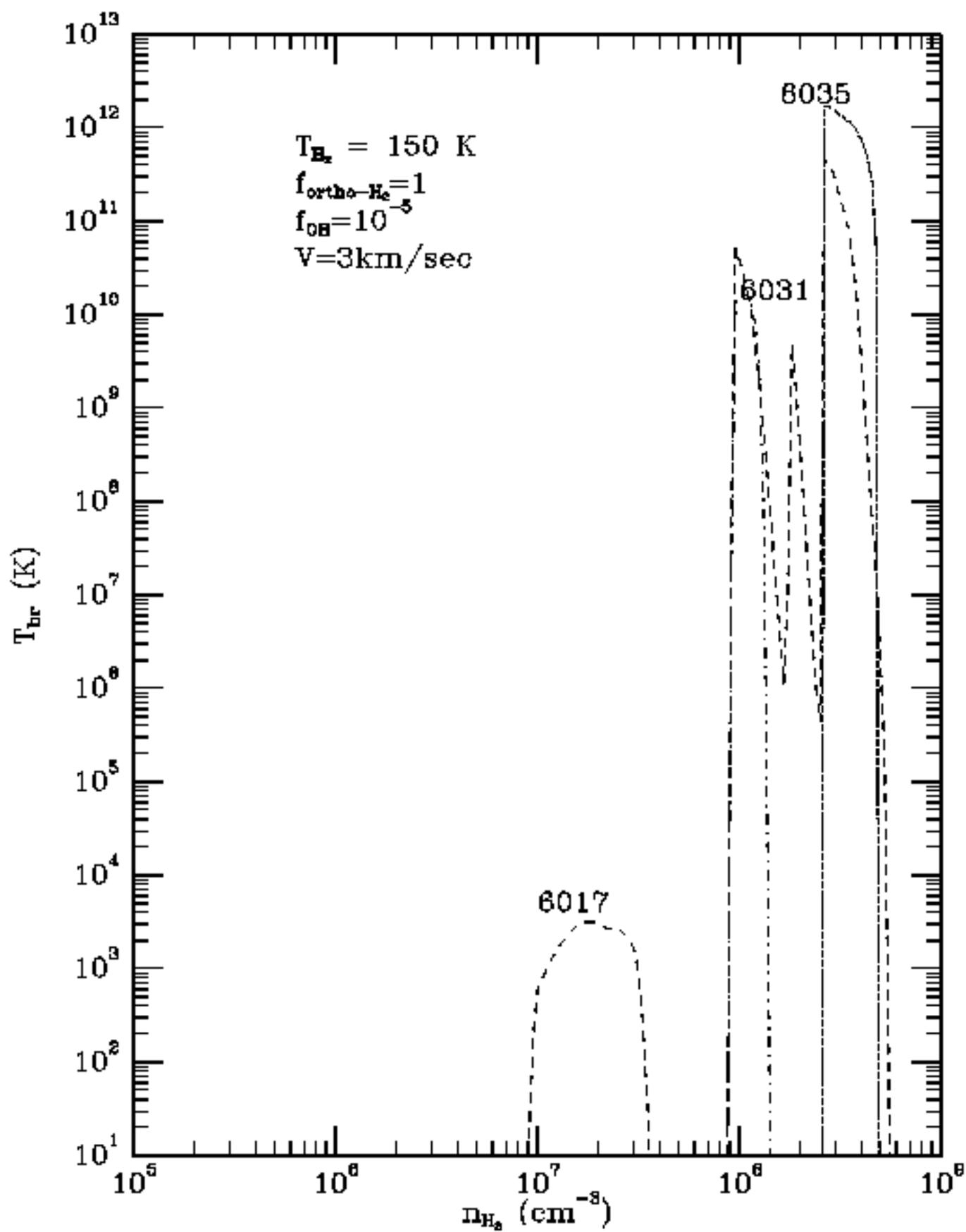
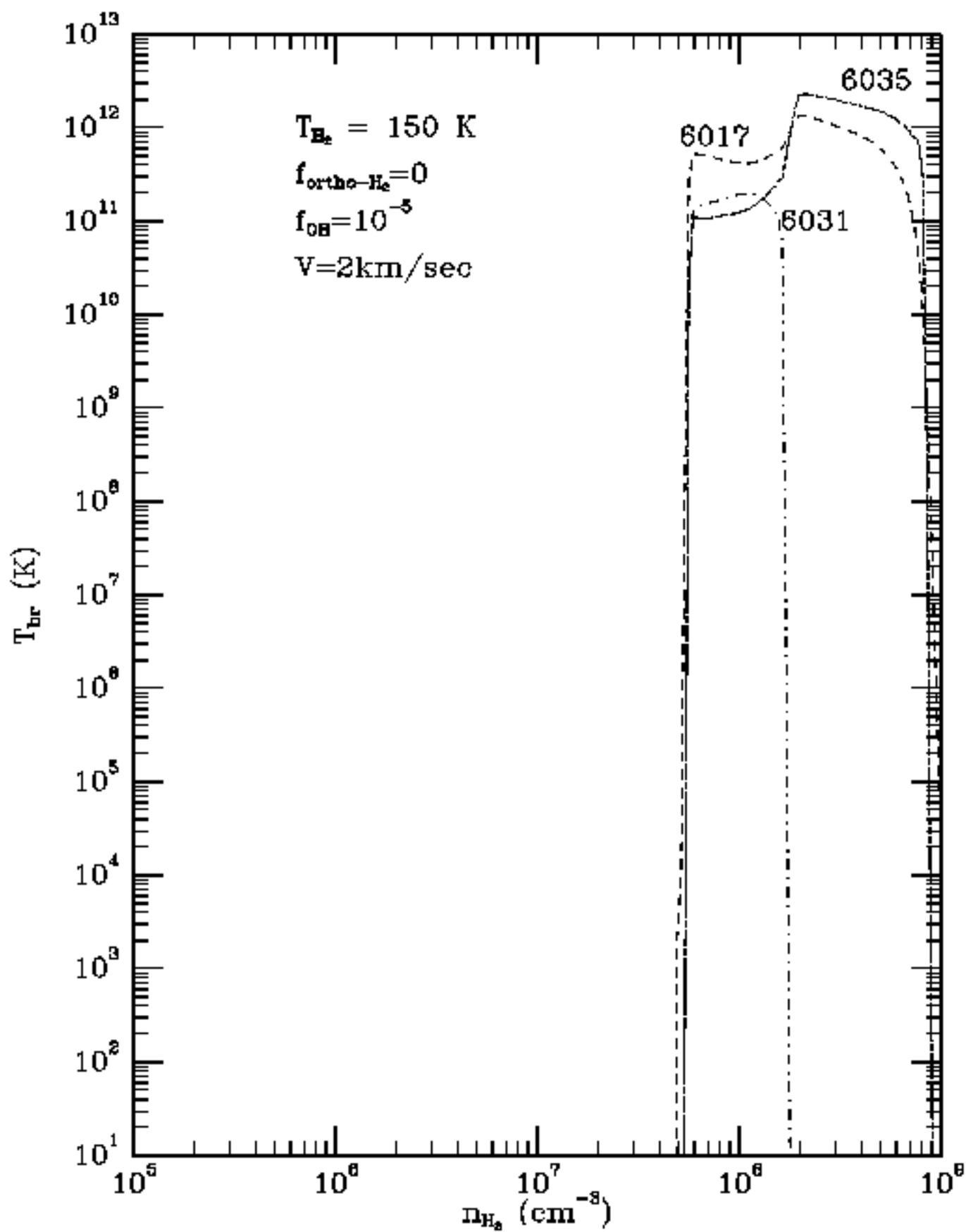
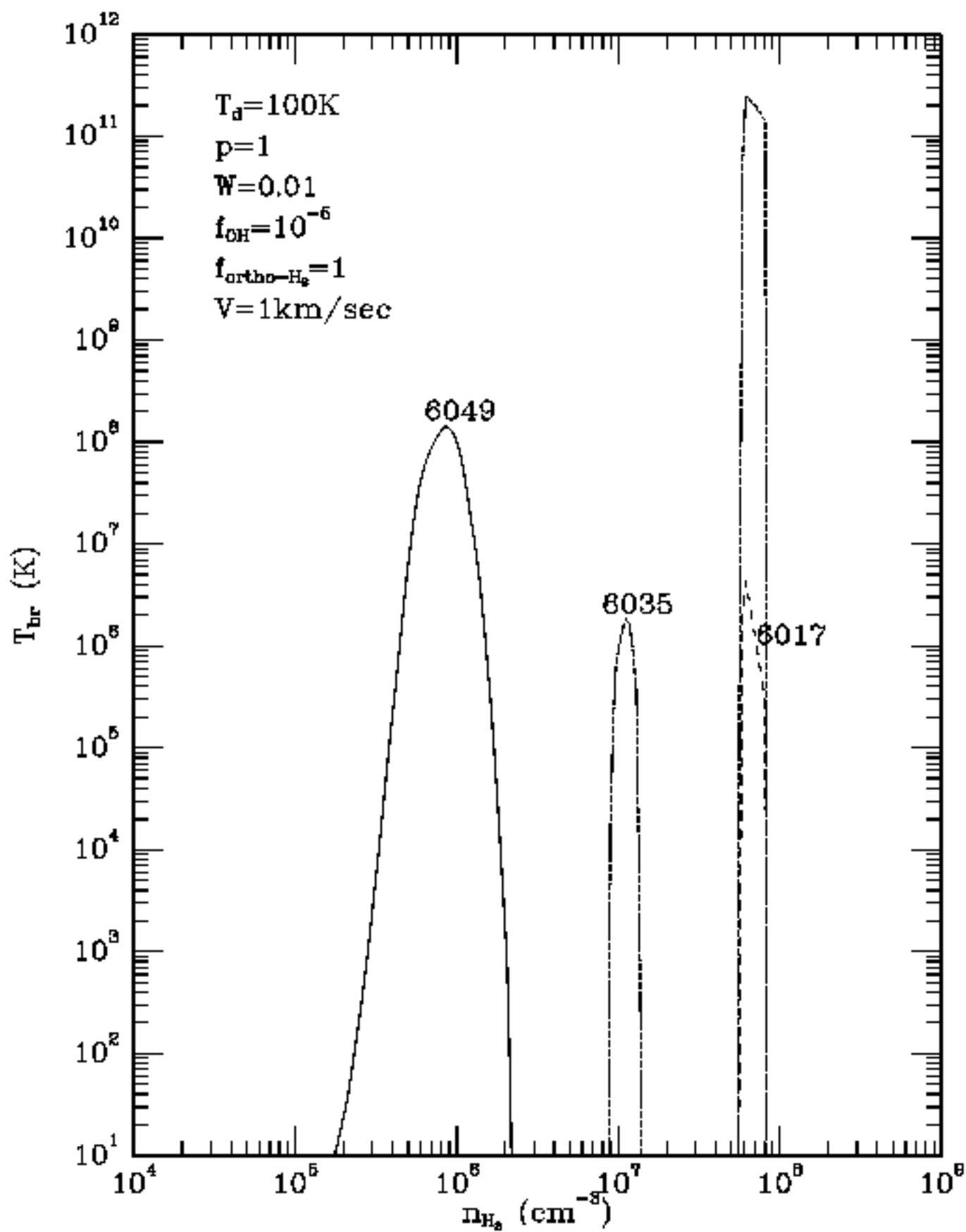
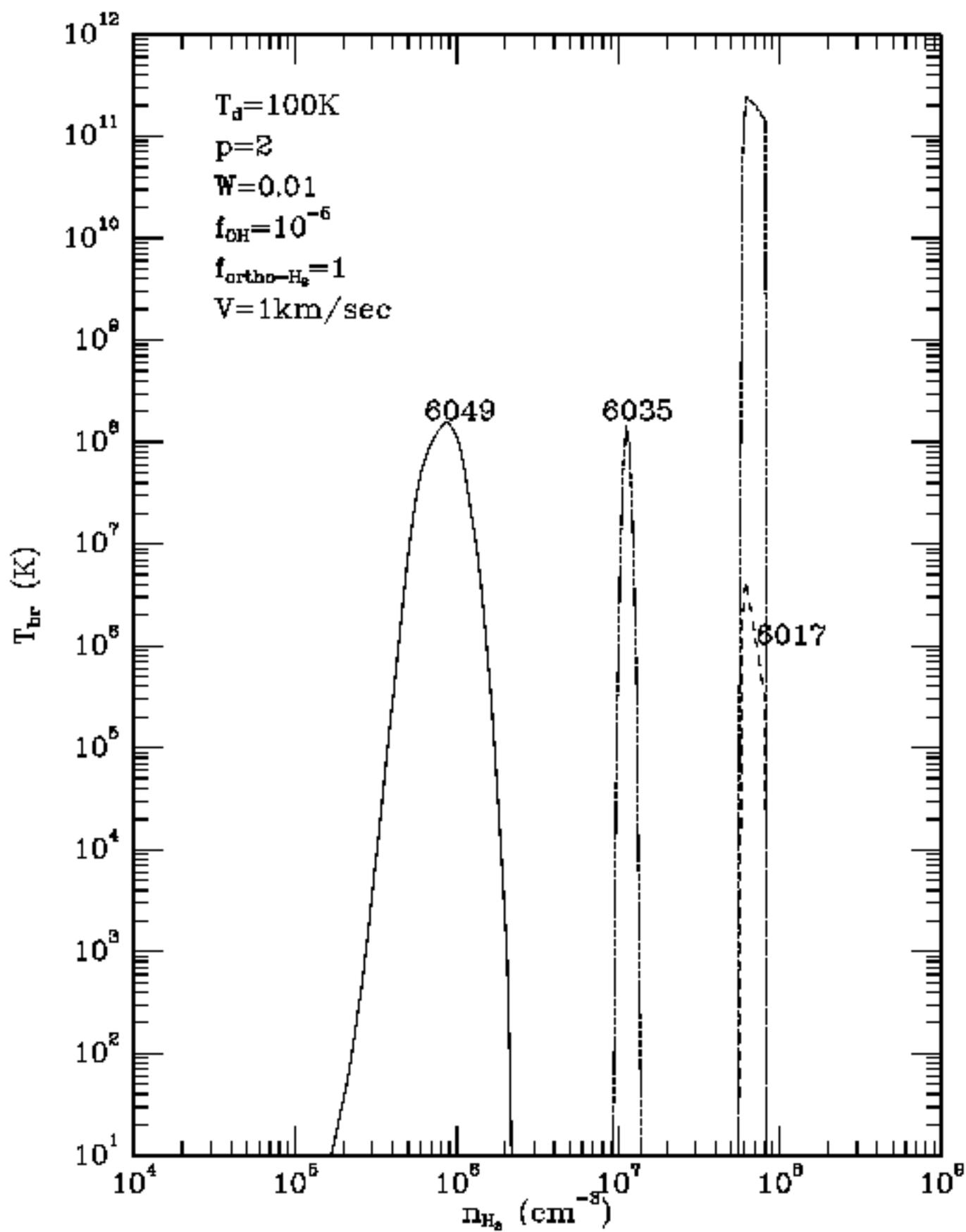
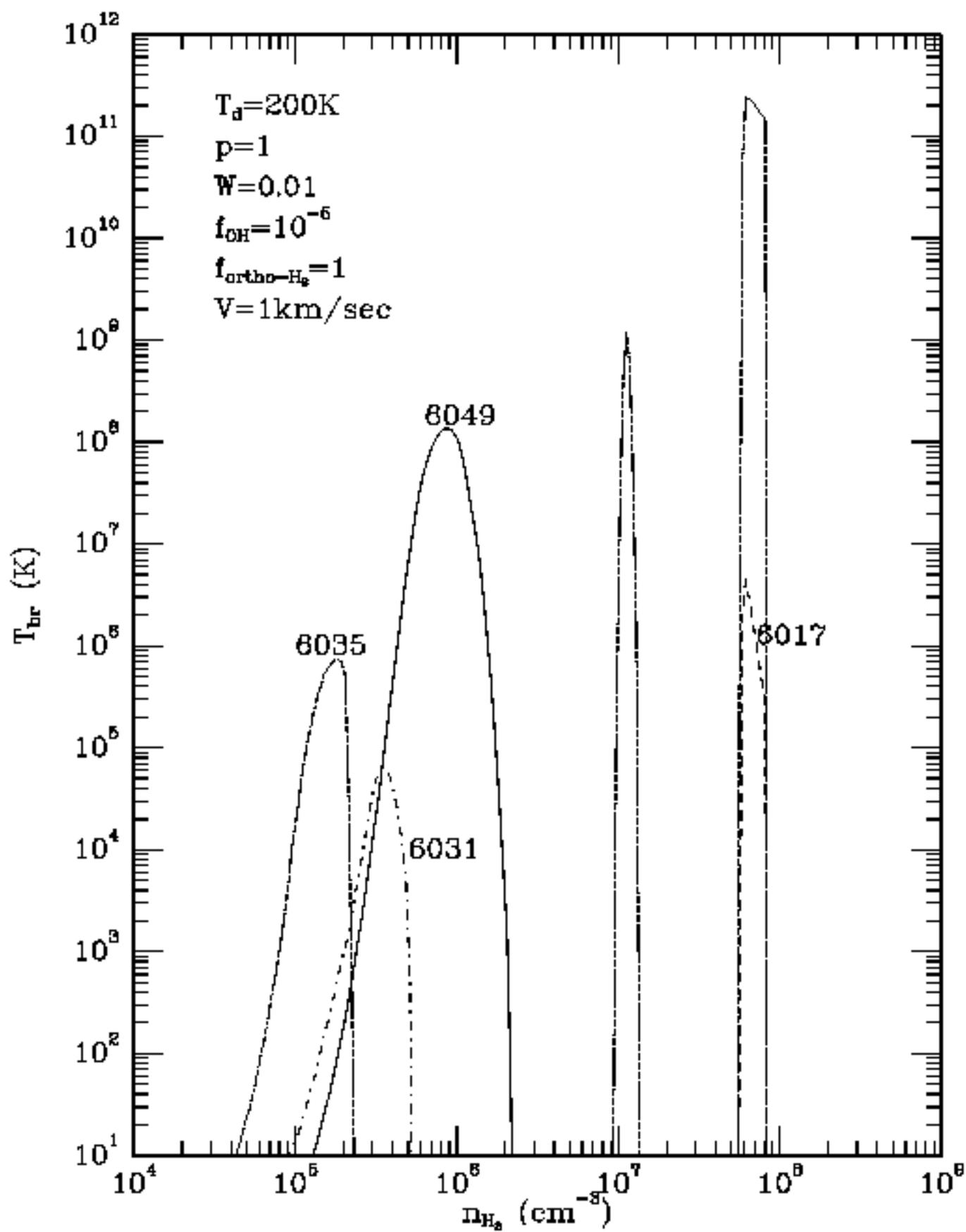
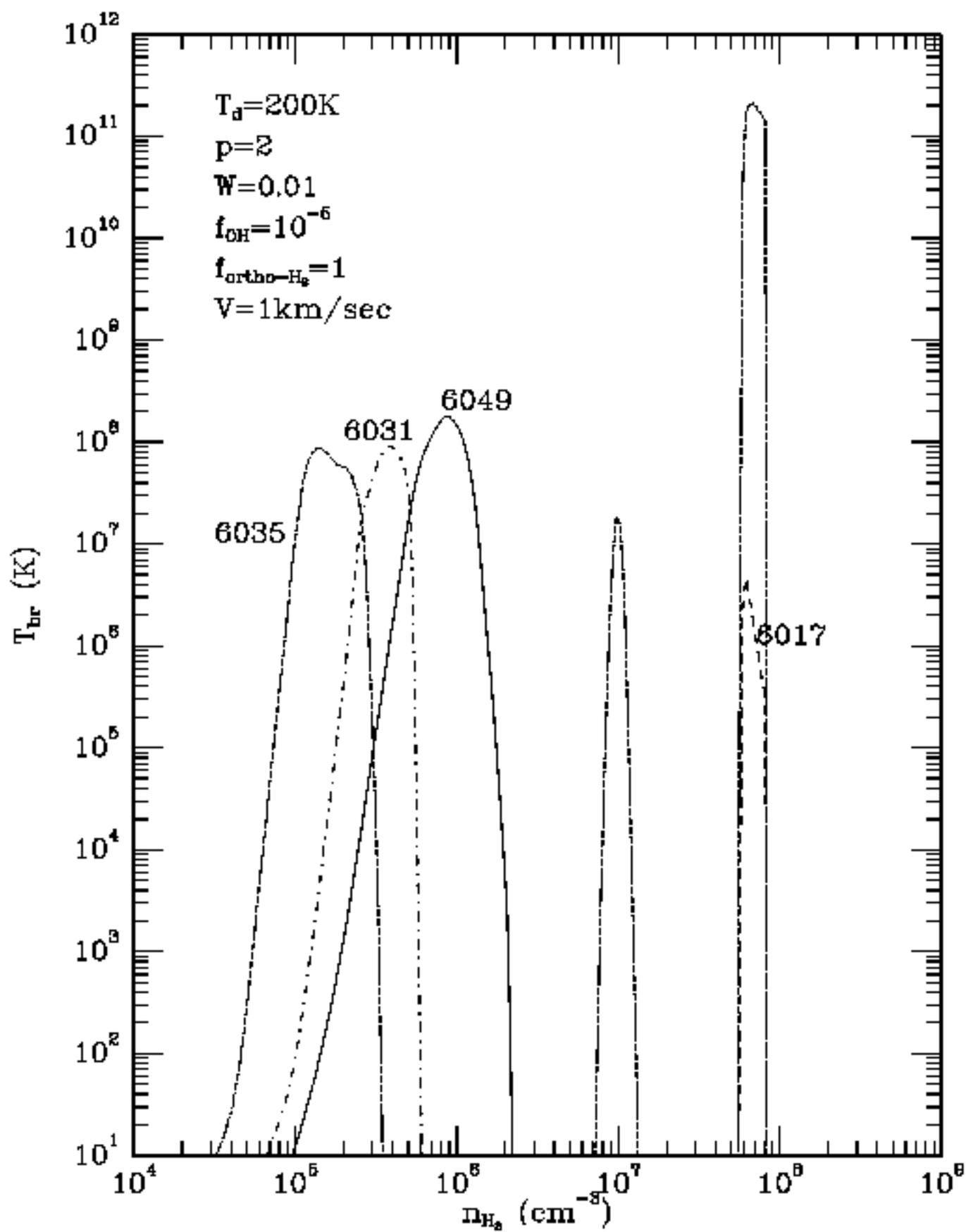
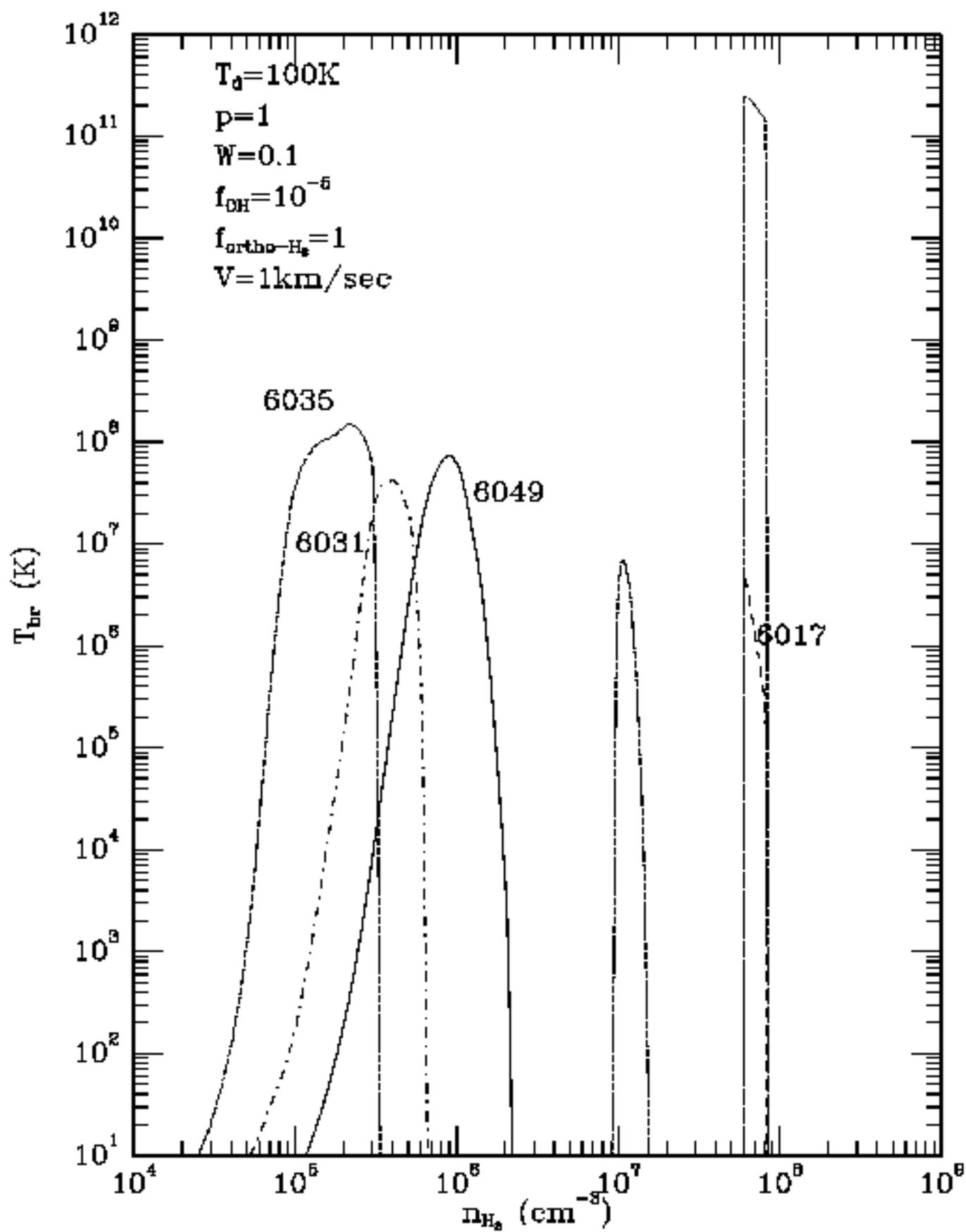
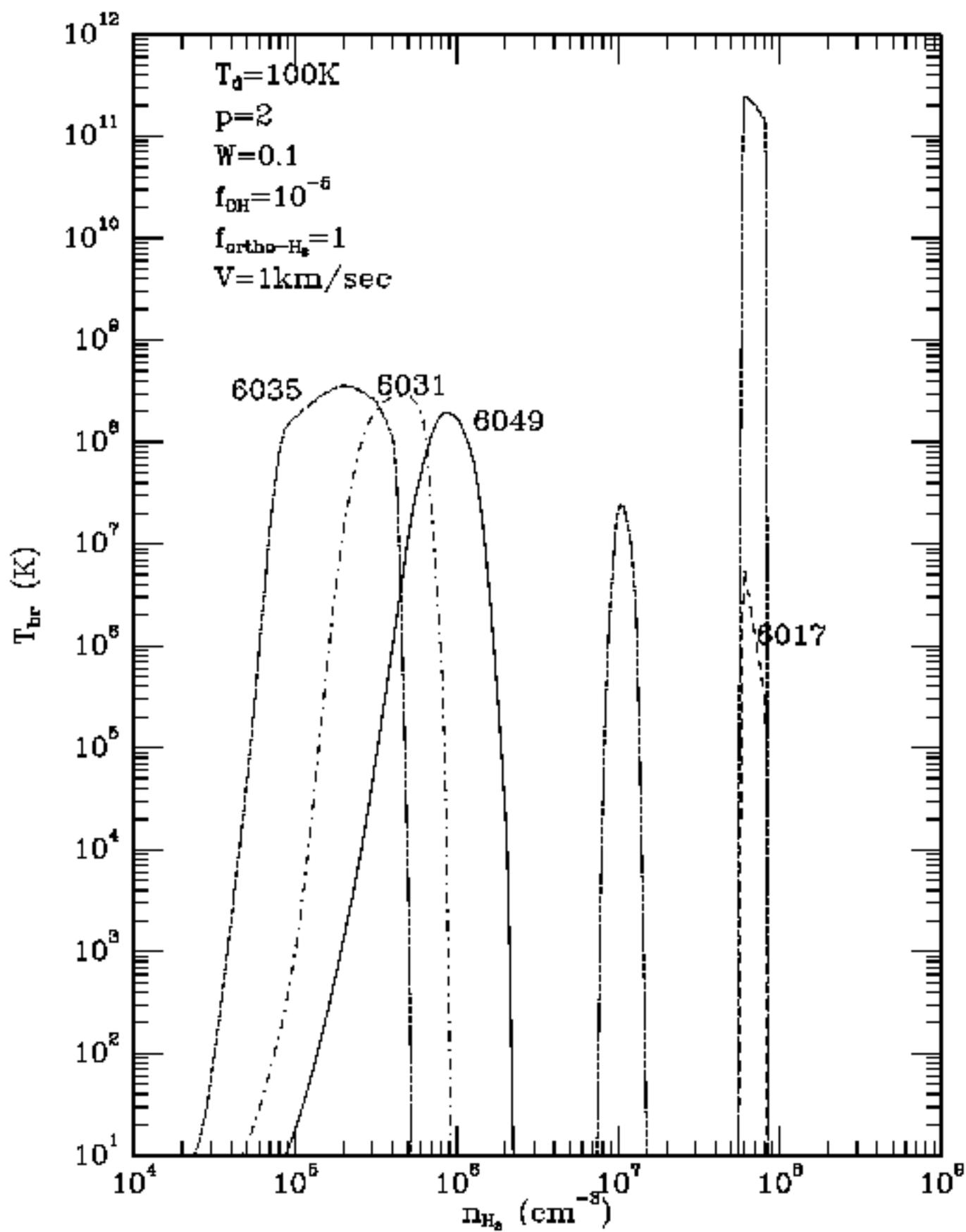
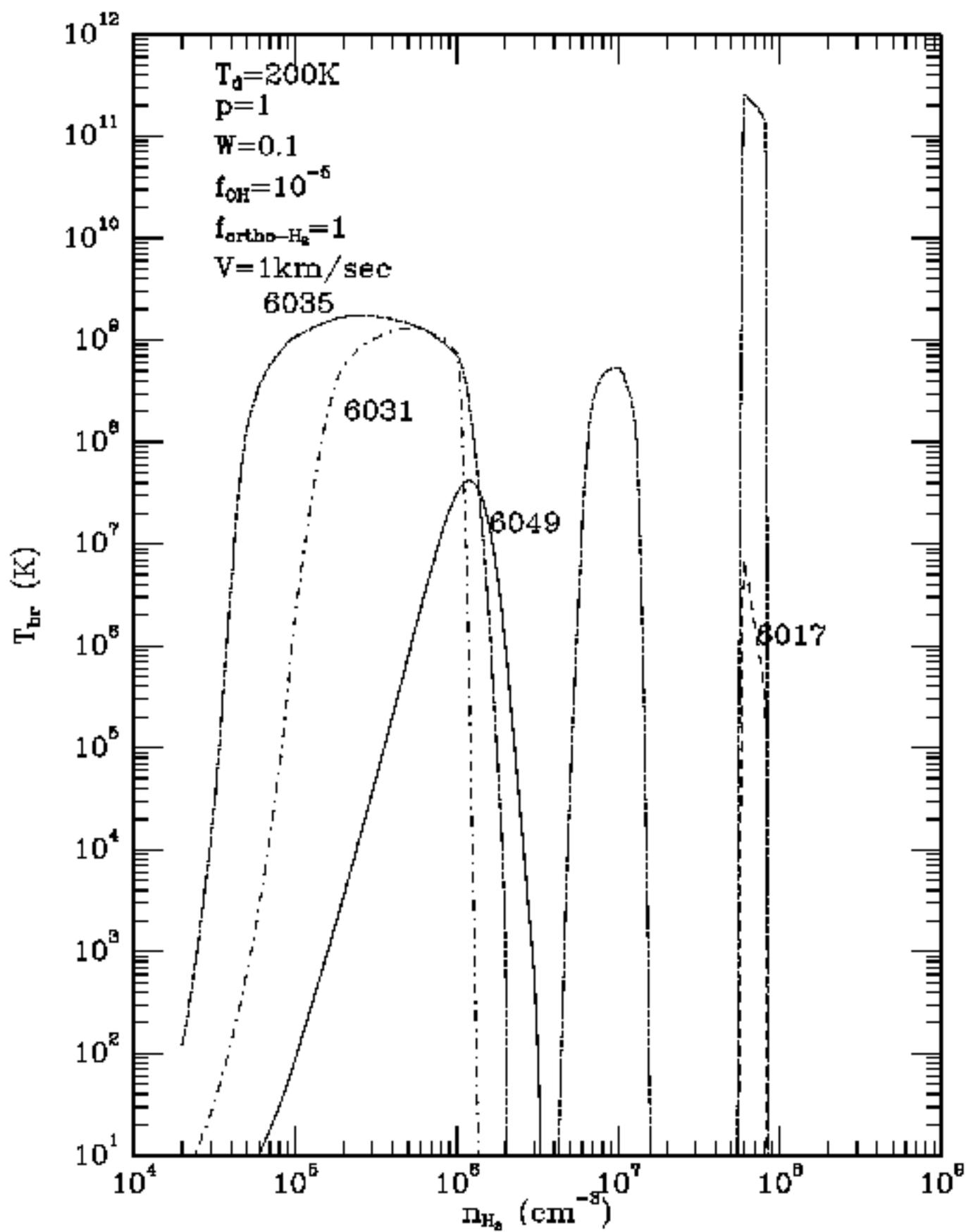
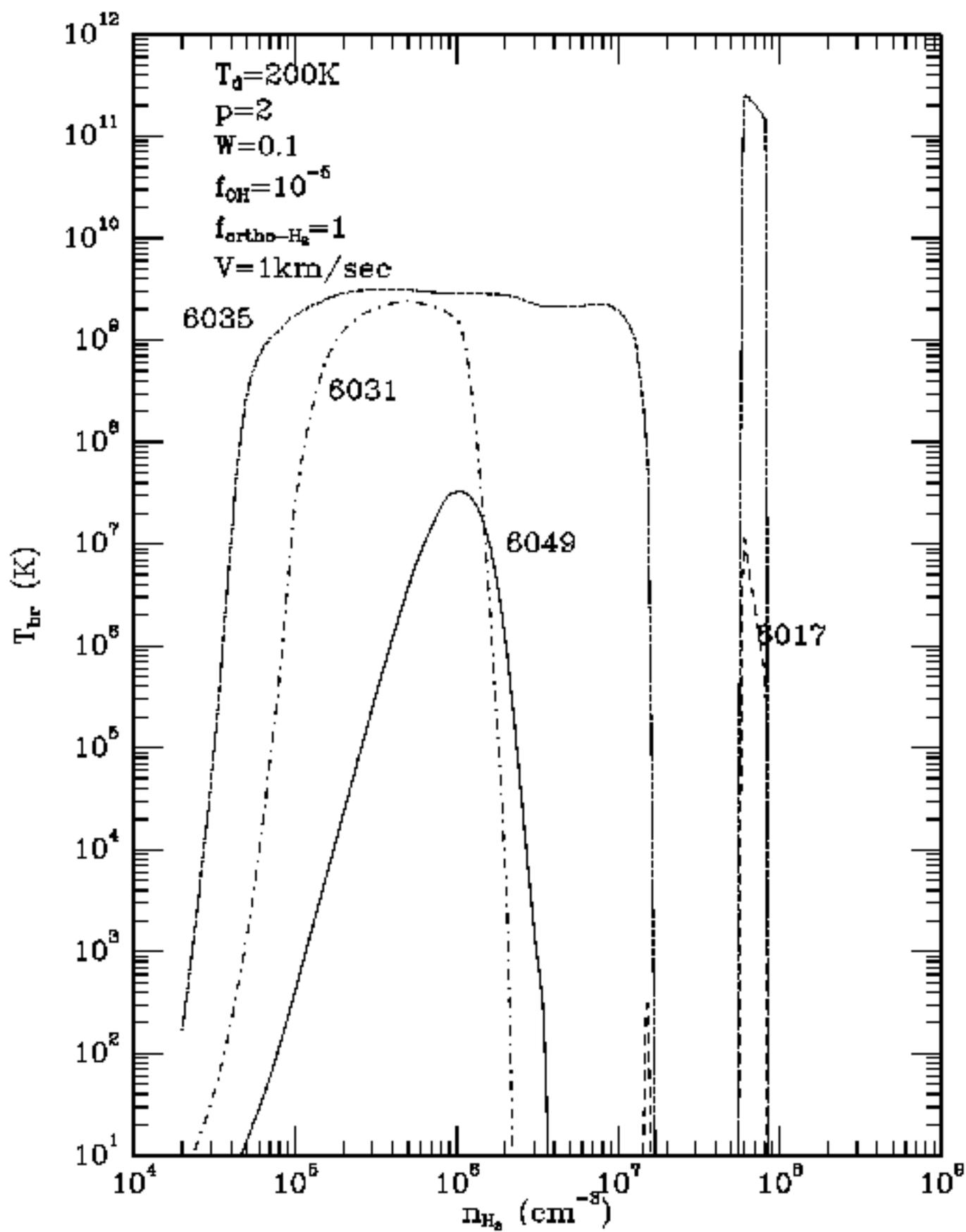
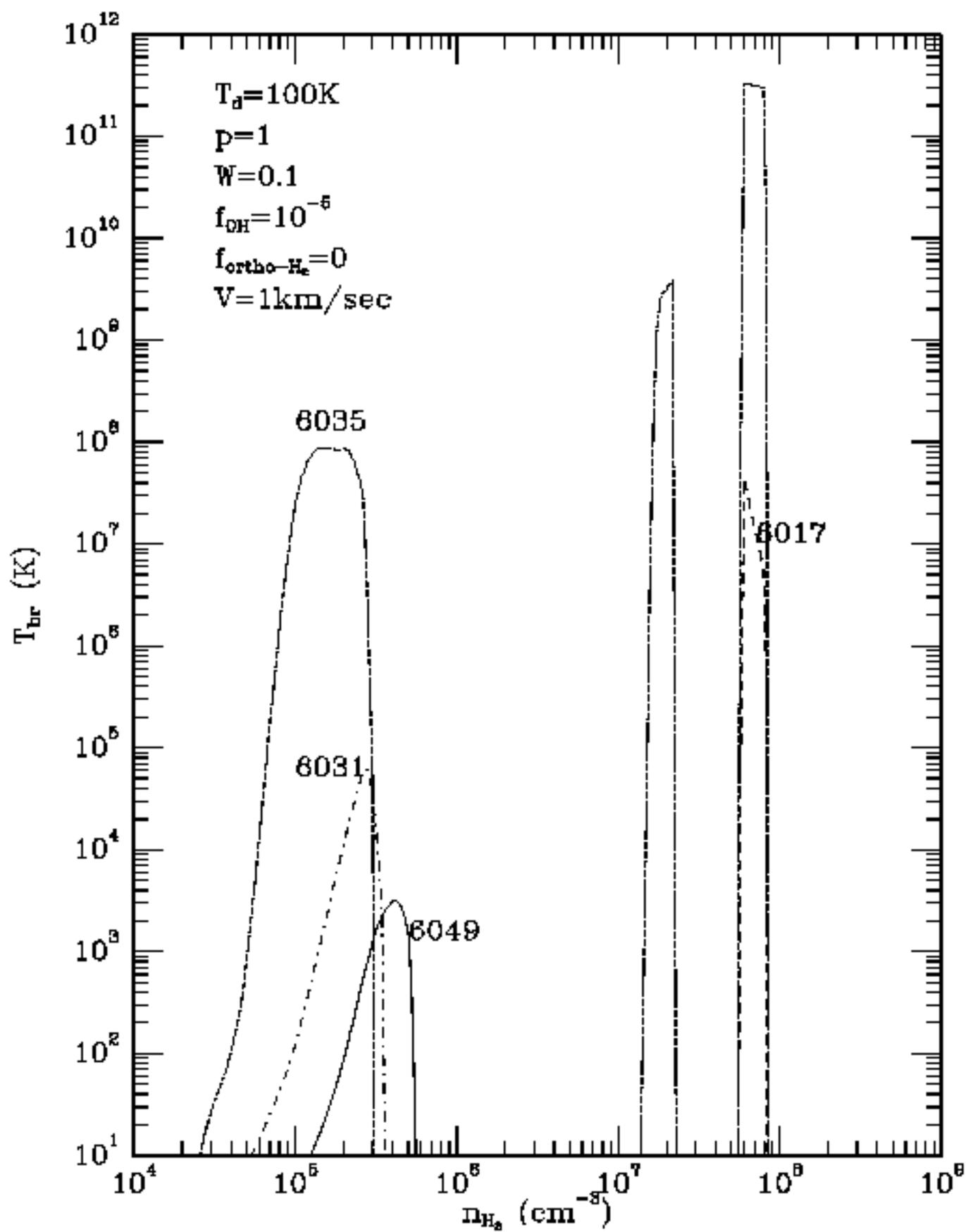
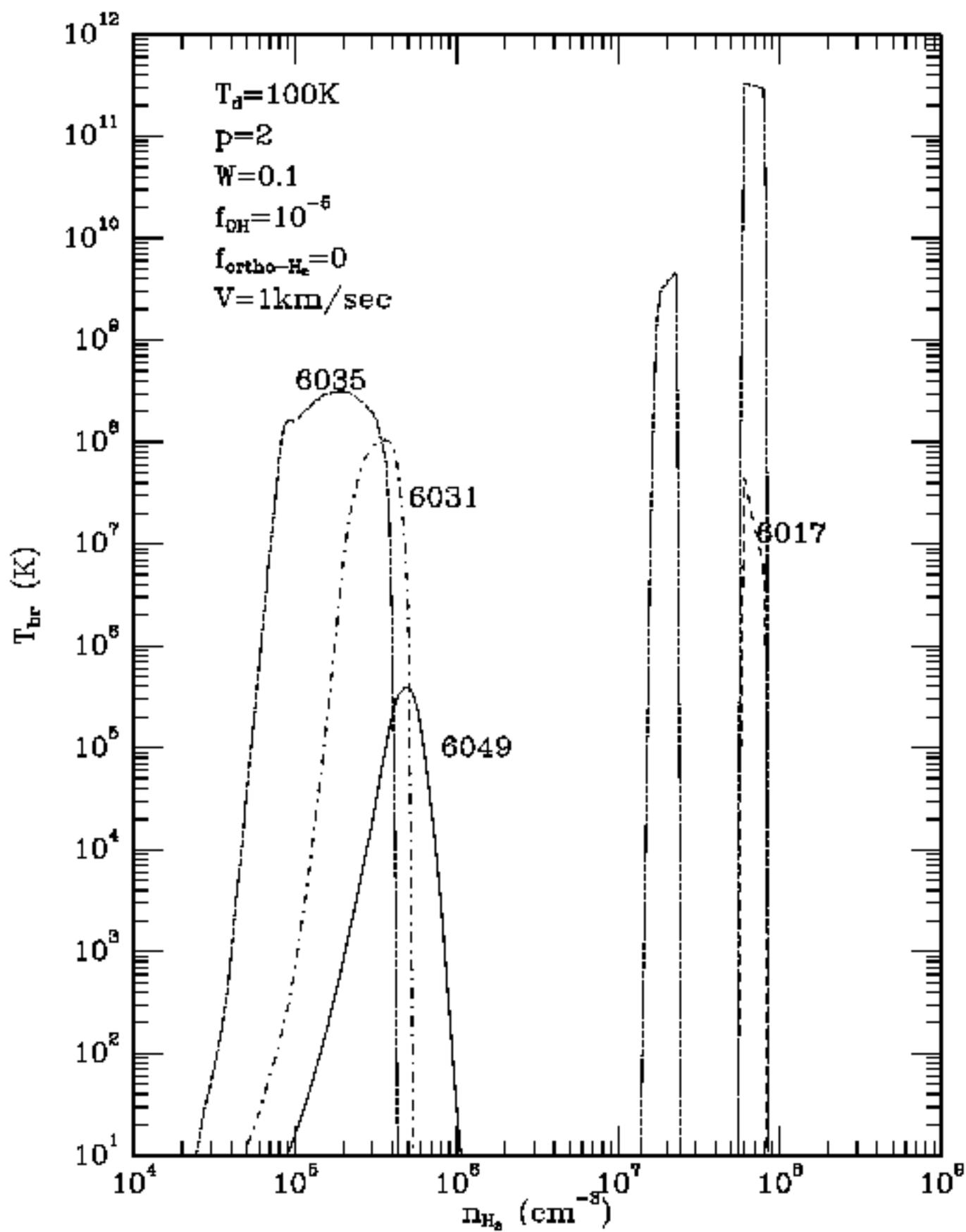
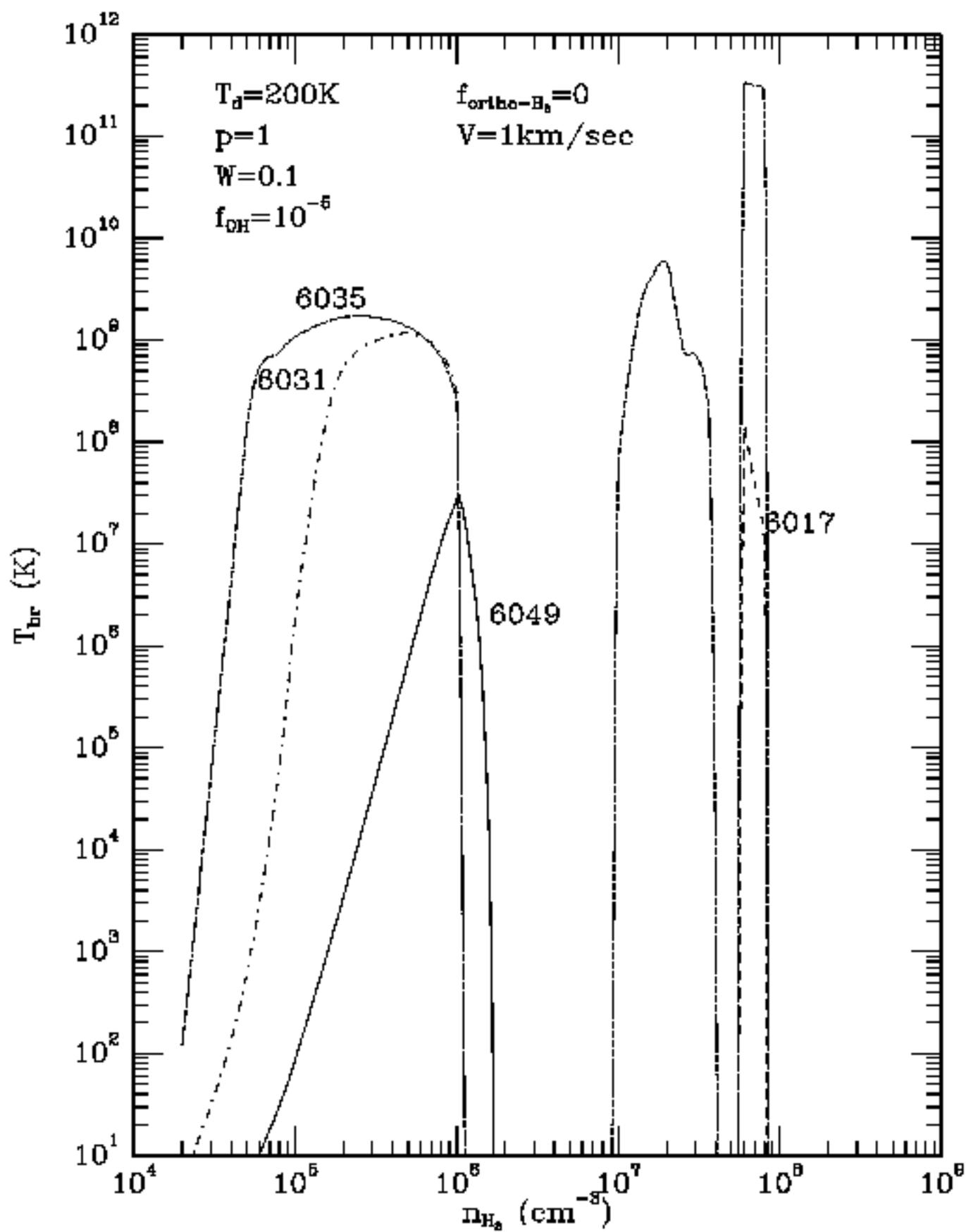
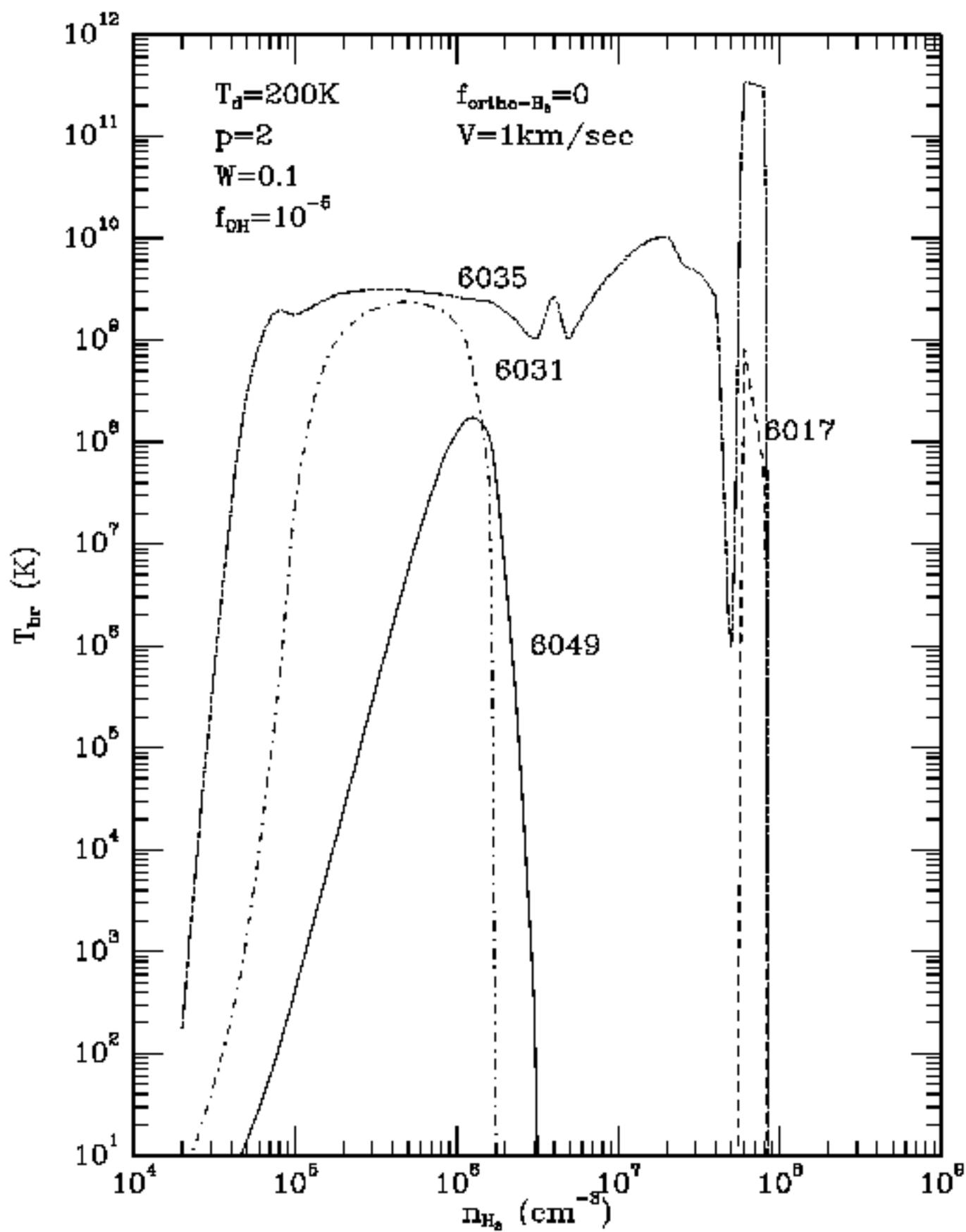
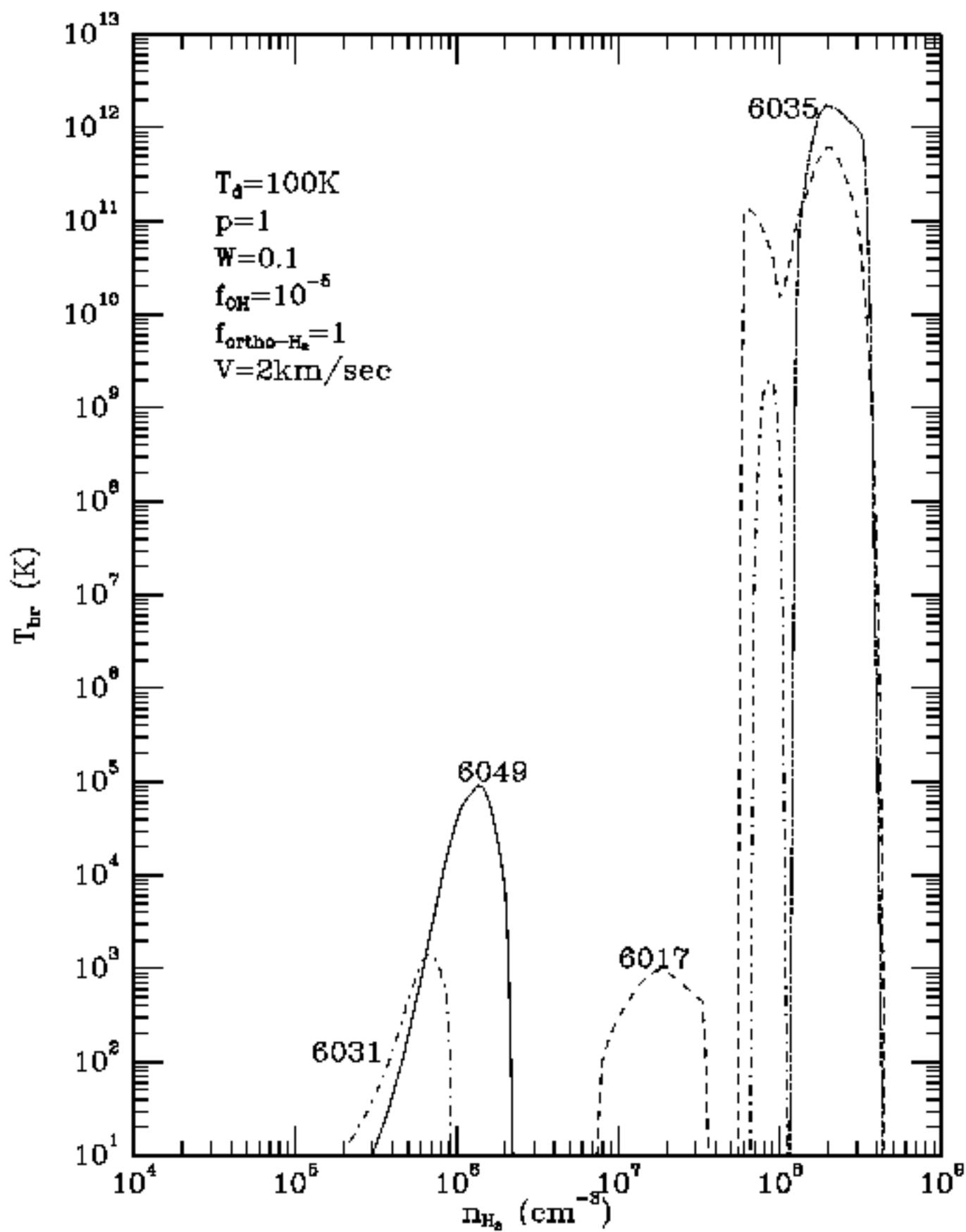
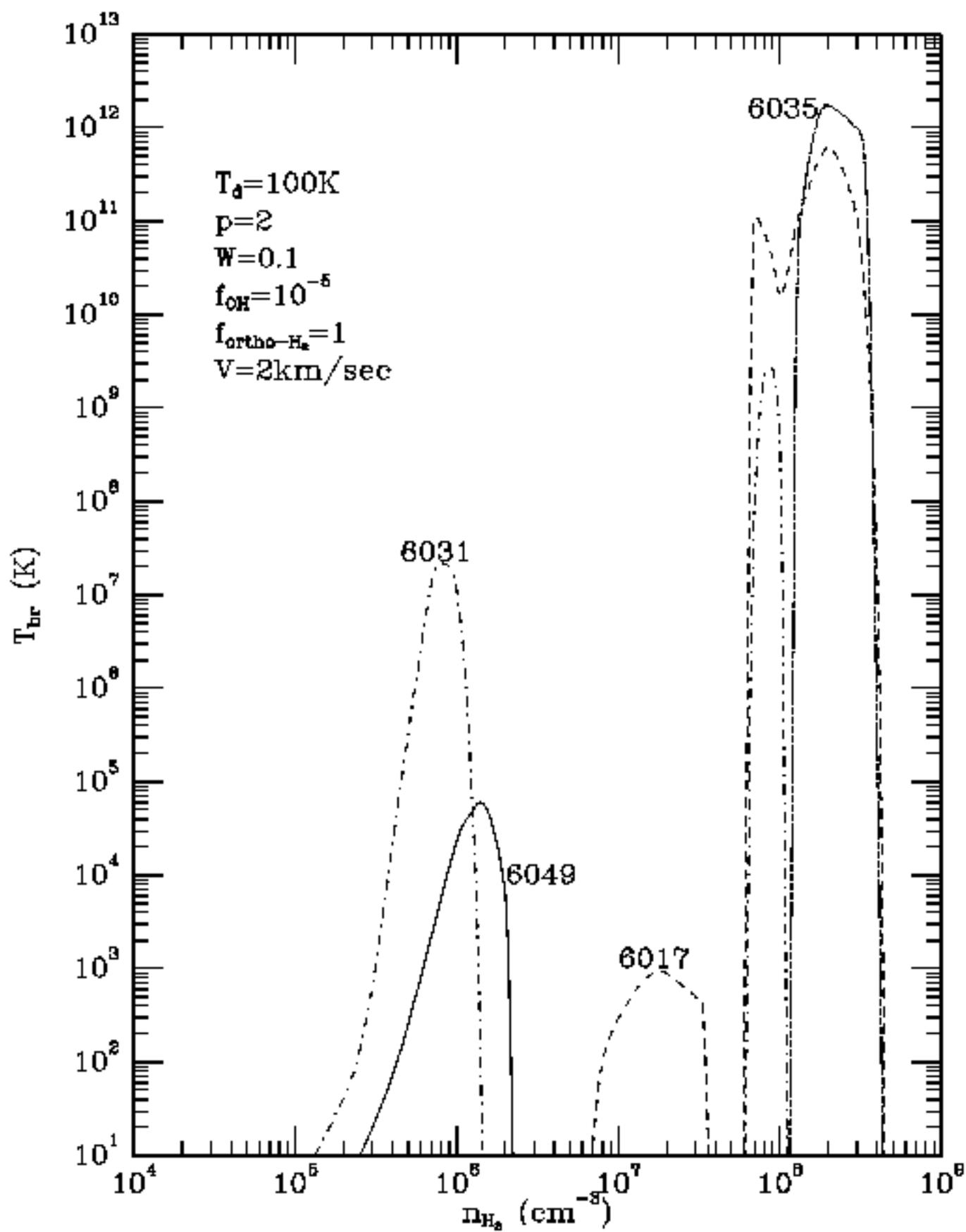
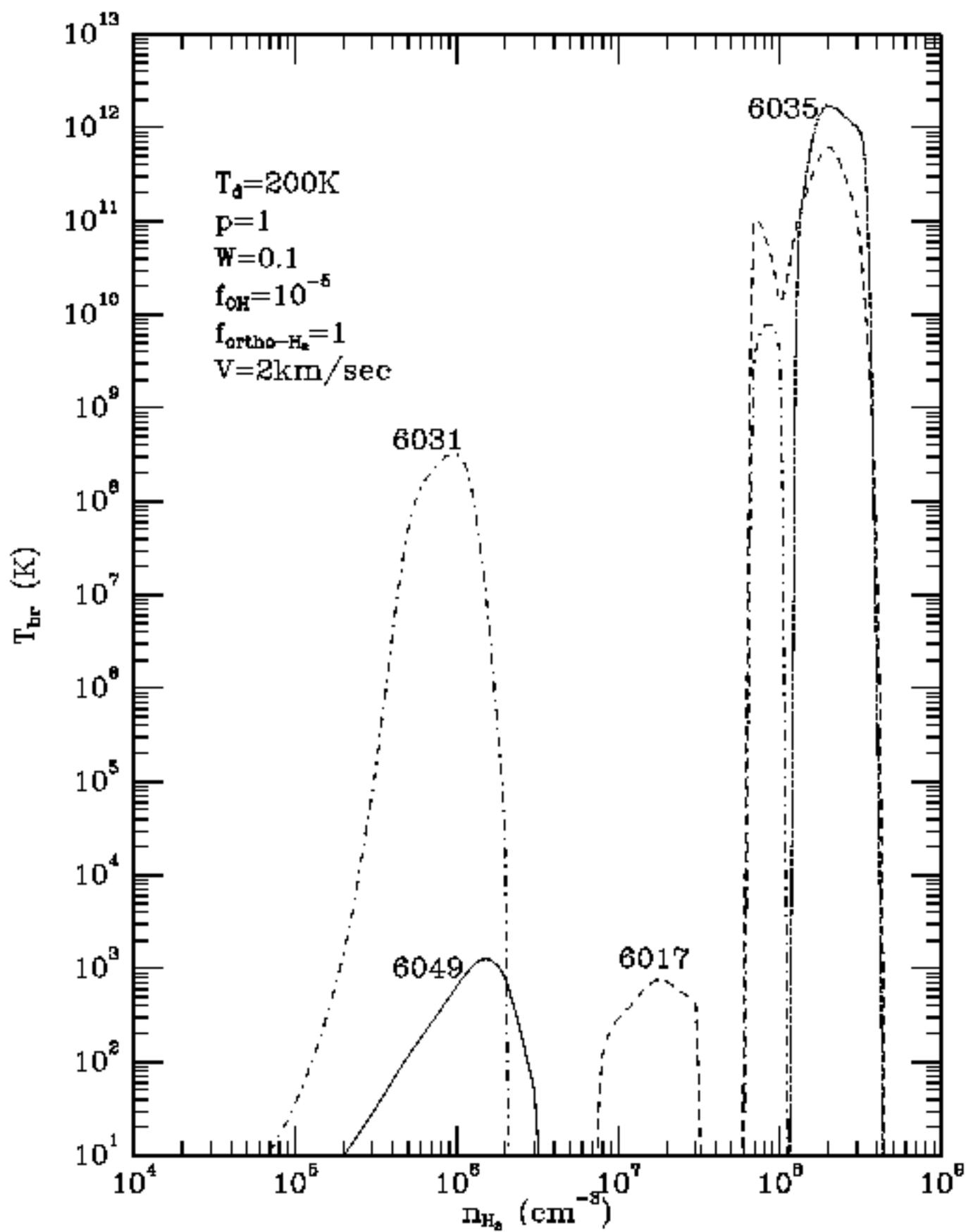
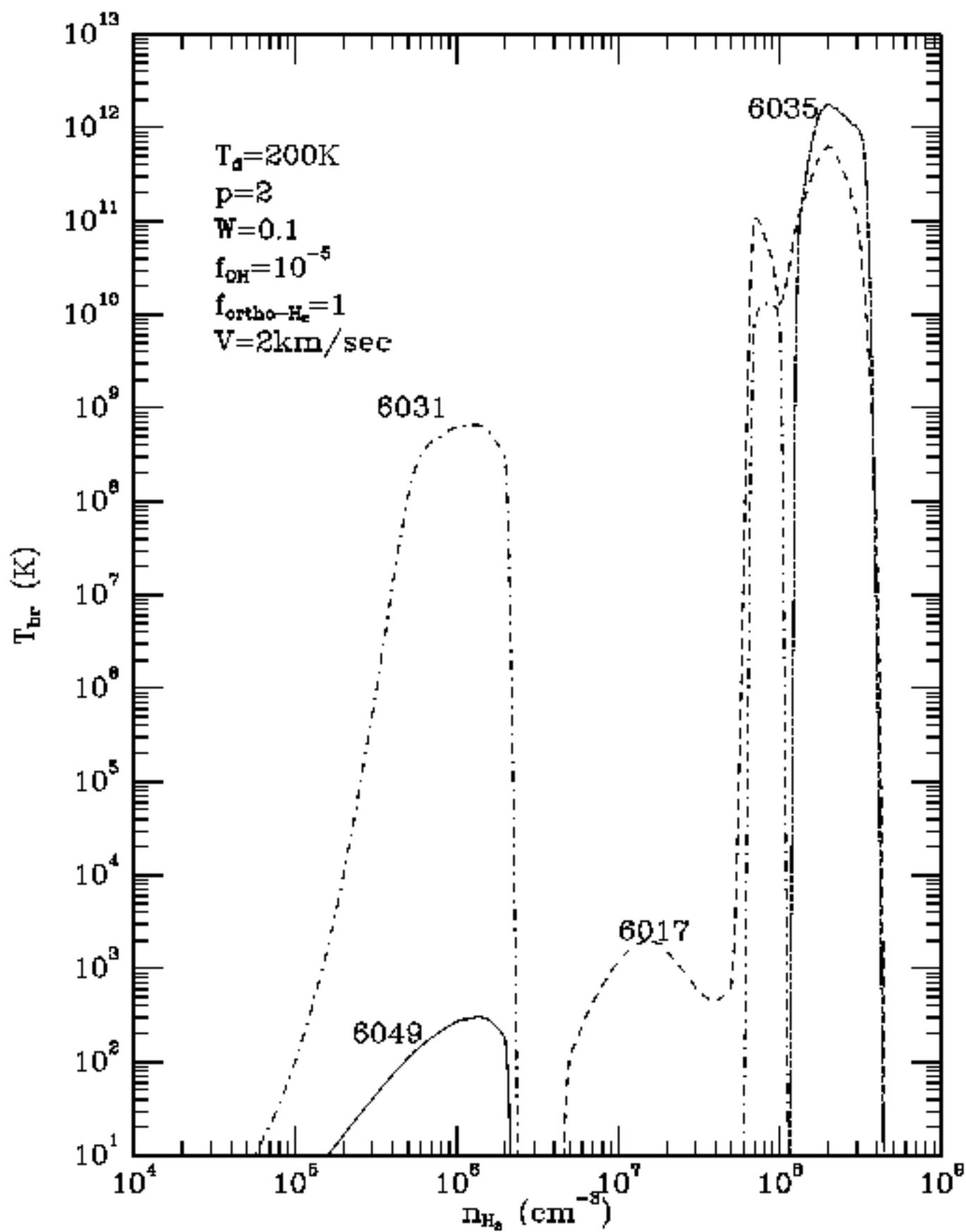
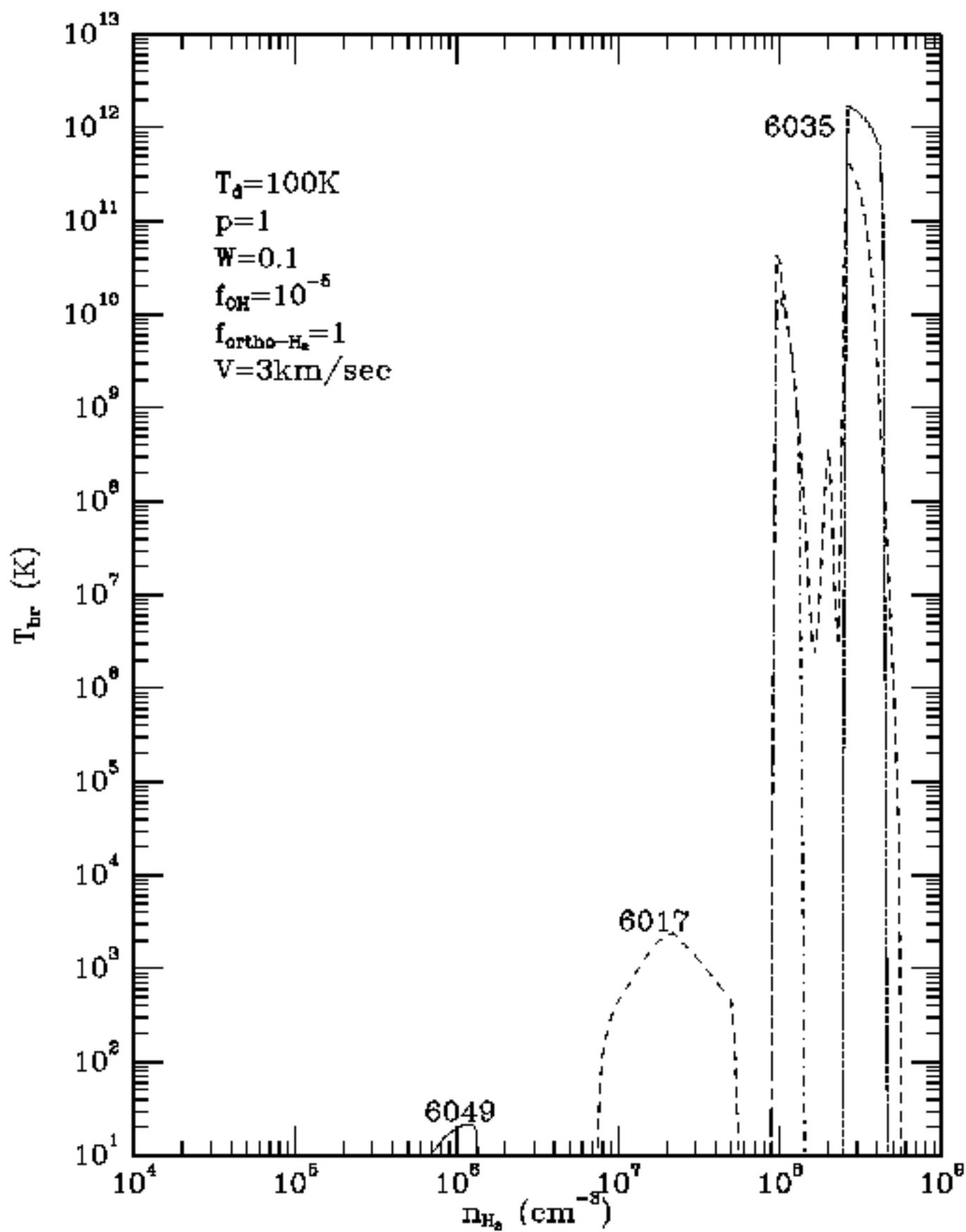
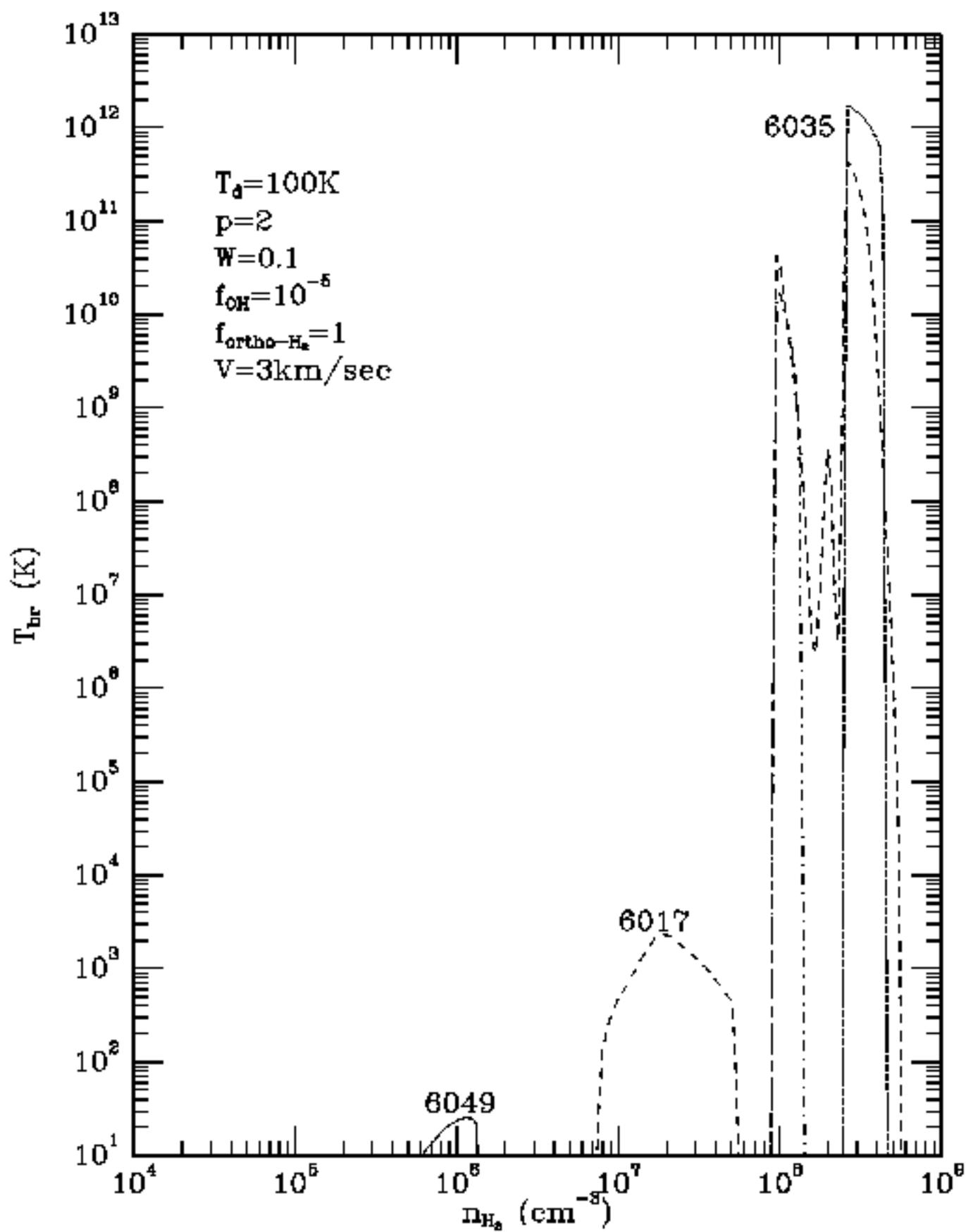
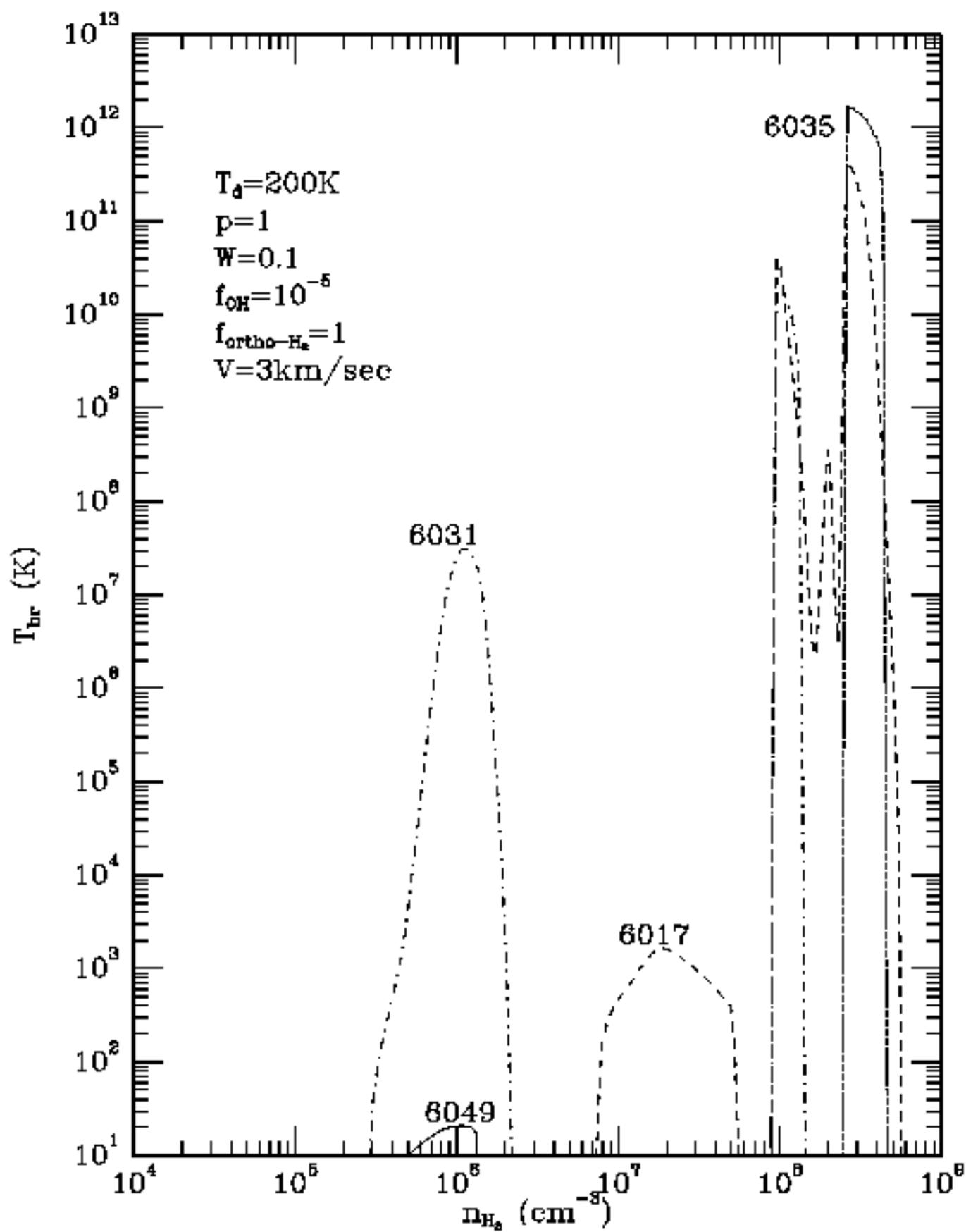
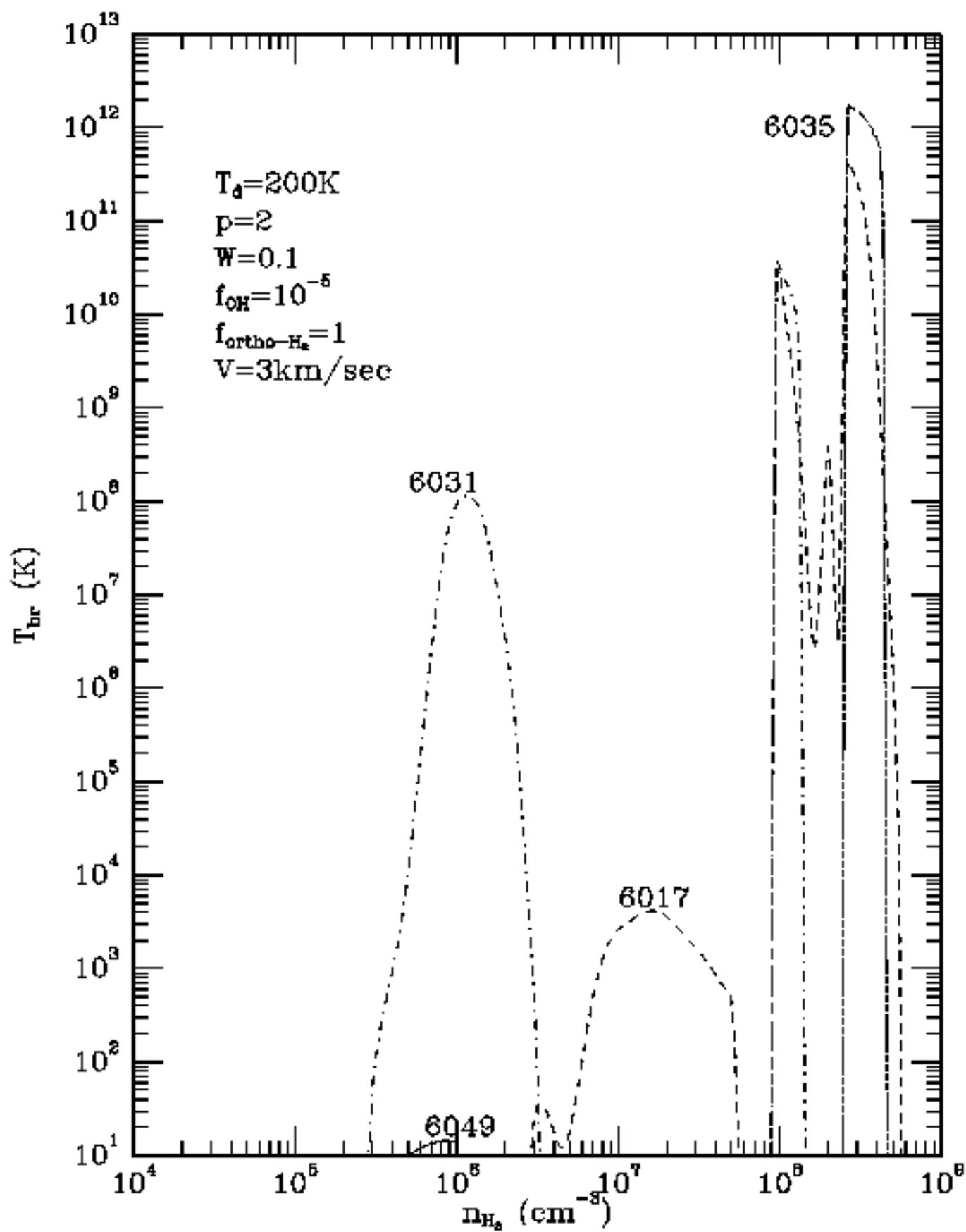
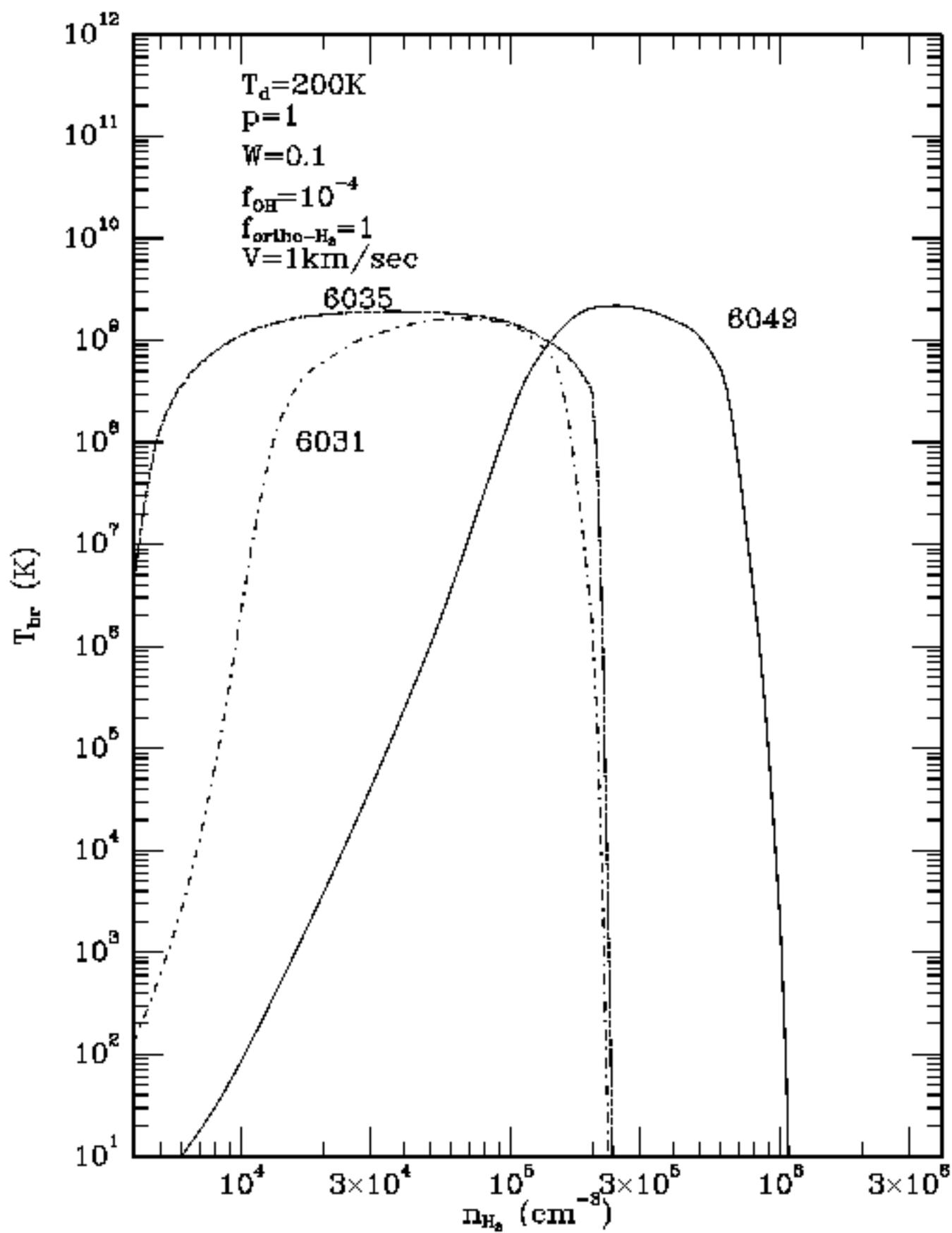
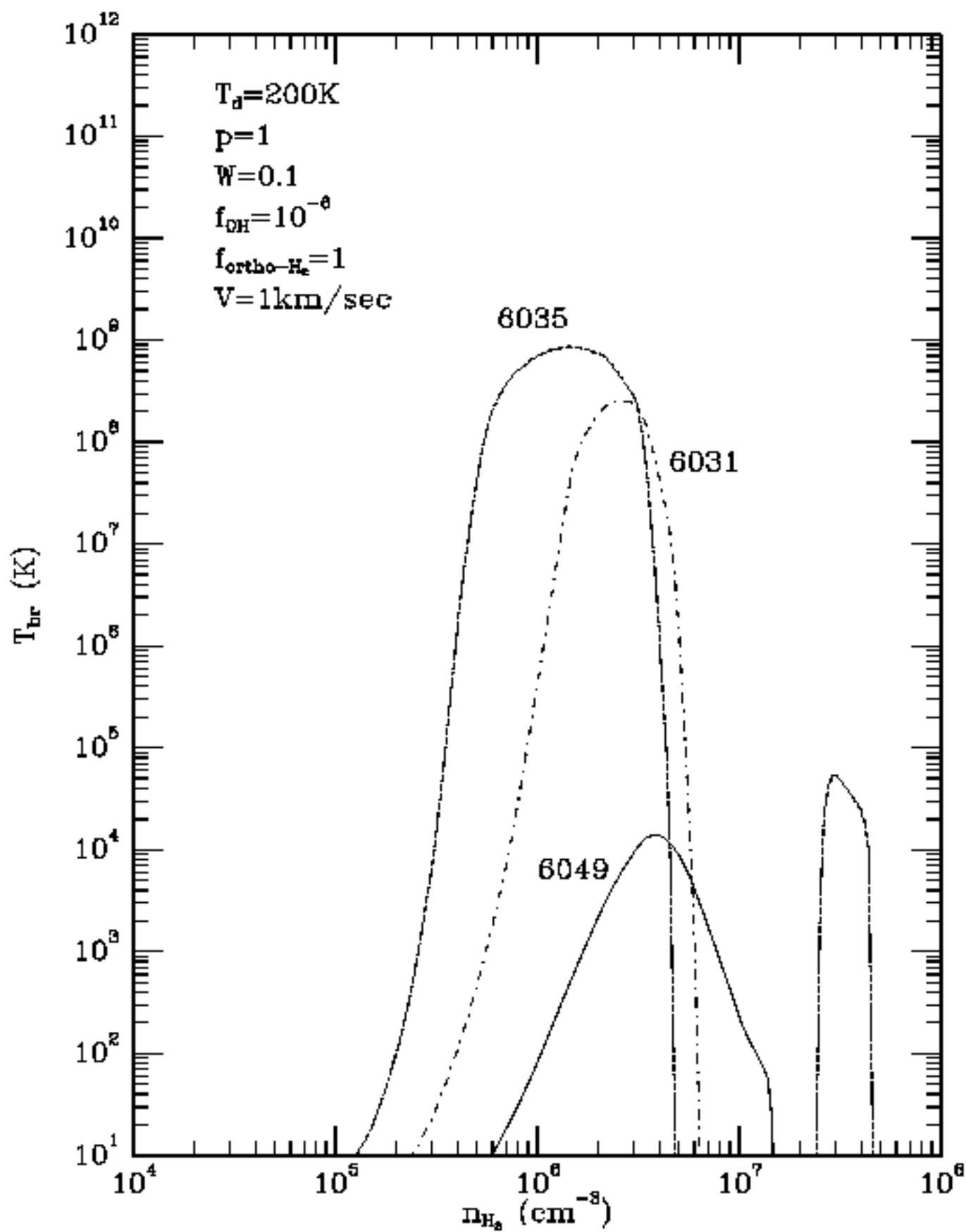
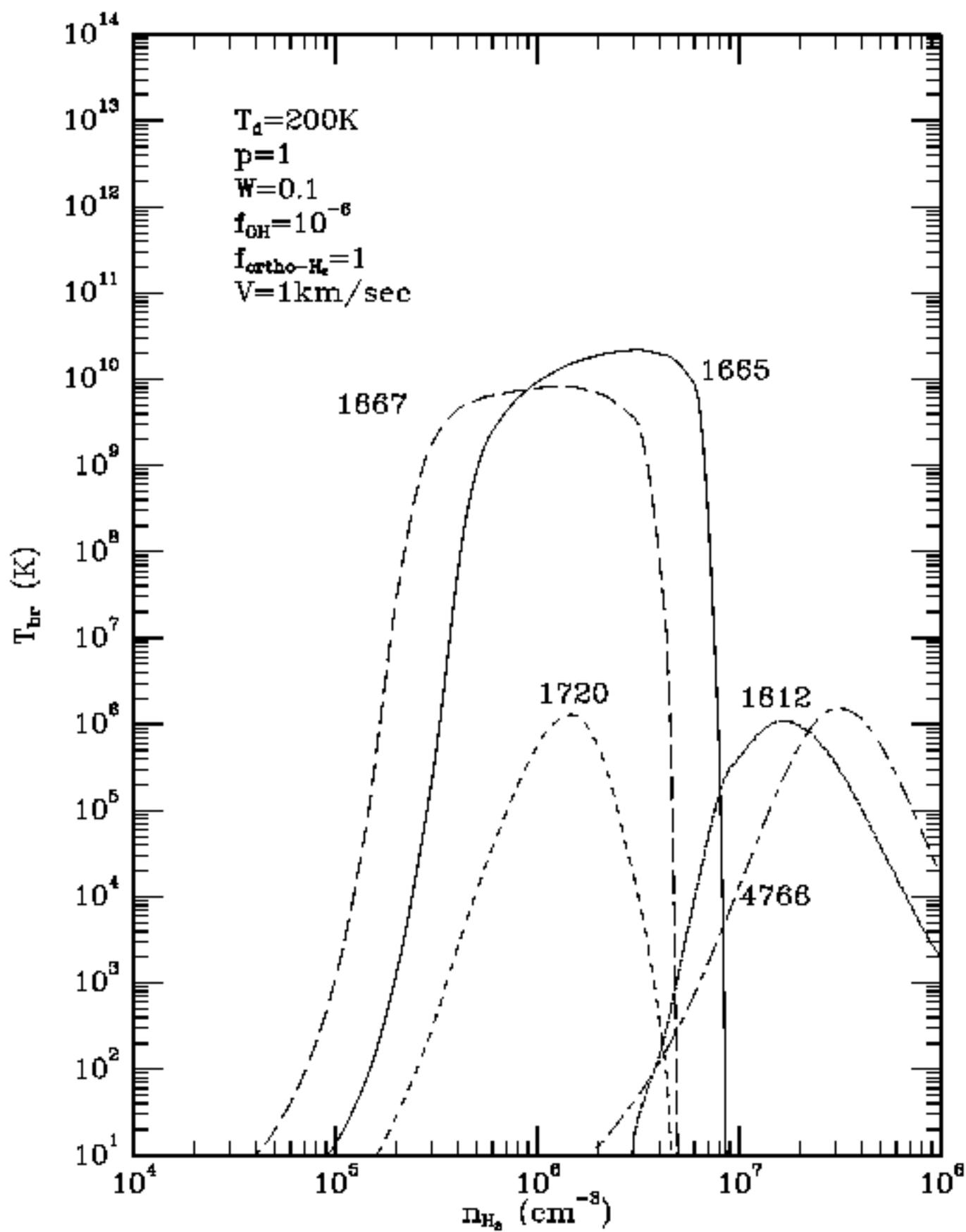
\bye